\documentclass[aps, prx, letter, superscriptaddress, twocolumn, amsfonts, amssymb, amsmath, reprint, showkeys, nofootinbib, twoside]{revtex4-2}
\usepackage[english]{babel}
\usepackage[sort&compress]{natbib}
\usepackage[utf8]{inputenc}
\usepackage[colorinlistoftodos, color=green!40, prependcaption]{todonotes}
\usepackage{amsthm}
\usepackage{mathtools}
\usepackage{physics}
\usepackage[version=4]{mhchem}
\usepackage{xcolor}
\usepackage{graphicx}
\usepackage{adjustbox}
\usepackage{placeins}
\usepackage[T1]{fontenc}
\usepackage{lipsum}
\usepackage{csquotes}
\usepackage{booktabs}
\usepackage[hypertexnames=false]{hyperref}
\hypersetup{colorlinks=true,citecolor=blue,linkcolor=blue,urlcolor=black}
\usepackage{siunitx}

\renewcommand{\selectlanguage}[1]{}

\def\be{\begin{equation}}
\def\ee{\end{equation}}
\def\bmu{\begin{multline}}
\def\bea{\begin{eqnarray}}
\def\eea{\end{eqnarray}}

\newcommand{\review}[1]{{{#1}}}

\begin{document}

\title{Nonequilibrium protein complexes as molecular automata}
\author{Jan Kocka}
\address{Department of Physics and Astronomy, University College London, London WC1E 6BT, United Kingdom}
\author{Kabir Husain}
\thanks{These authors contributed equally. Email: kabir.husain@ucl.ac.uk, j.agudo-canalejo@ucl.ac.uk}
\address{Department of Physics and Astronomy, University College London, London WC1E 6BT, United Kingdom}
\address{Laboratory for Molecular Cell Biology, University College London, London WC1E 6BT, United Kingdom}
\author{Jaime Agudo-Canalejo}
\thanks{These authors contributed equally. Email: kabir.husain@ucl.ac.uk, j.agudo-canalejo@ucl.ac.uk}
\address{Department of Physics and Astronomy, University College London, London WC1E 6BT, United Kingdom}

\begin{abstract}
	Biology stores information and computes at the molecular scale, yet the ways in which it does so are often distinct from human-engineered computers. Mapping biological computation onto architectures familiar to computer science remains an outstanding challenge. Here, inspired by Crick’s proposal for molecular memory, we analyse a thermodynamically-consistent model of a protein complex subject to driven, nonequilibrium enzymatic reactions. In the strongly driven limit, we find that the system maps onto a stochastic, asynchronous variant of cellular automata, where each rule corresponds to a different set of enzymes being present. We find a broad class of phenomena in these `molecular automata' that can be exploited for molecular computation, including error-tolerant memory via multistable attractors, and long transients that can be used as molecular stopwatches. By systematically enumerating all possible dynamical rules, we identify those that allow molecular automata to implement simple computational architectures such as finite-state machines. Overall, our results provide a framework for engineering synthetic molecular automata, and offer a route to building protein-based computation in living cells.
\end{abstract}

\maketitle

\section{Introduction}

Molecular computation in biology and biomimetic systems is enabled by multistability, with each stable configuration of the system representing a distinct logical state or memory \cite{hopfield1982neural}. In physical systems, these often correspond to multiple minima in an energy landscape, as in (spin) glasses \cite{RevModPhys.58.801}, self-folding origami \cite{PhysRevX.7.041070}, or multifarious self-assembly \cite{murugan2015multifarious}. A similar framework is invoked in biological systems, though these are often discussed in the language of attractors and fixed points in some underlying dynamical system, as in neural activity patterns \cite{khona2022attractor}, the Waddington landscape of development \cite{waddington2014strategy,raju2023theoretical}, or multistable fates in gene regulatory networks \cite{ptashne,ozbudak2004multistability}.

Beyond natural systems, there is a growing interest in building synthetic multistable circuits at the molecular scale. However, outside of a few instances, synthetic implementations are centred around DNA: either in conjunction with DNA-binding proteins, as in circuits made of synthetic transcription factors \cite{zhu2022synthetic} or site-specific DNA recombinases \cite{roquet2016synthetic}, or alone as in strand-displacement circuits assembled in-vitro \cite{qian2011scaling}. In contrast, natural biological signal processing often involves large, multi-protein complexes that `compute' via post-translational modifications \cite{chen2021programmable}. The design rules that underlie the computational capabilities of such protein circuits remain broadly unknown.

In 1984, Francis Crick proposed a particular mechanism by which protein complexes can harbour long-term memory despite stochasticity and molecular turn-over \cite{crick1984neurobiology}. The basis of his construction was a dimer of monomers that can each be activated (e.g., phosphorylated) by an enzyme. If, as he argued, the enzyme only acts on a monomer when the other monomer is already activated, then the dimer is capable of storing two distinct steady states: both monomers unactivated, or both activated. Crucially, as in an error-correcting code \cite{hamming1950error}, the fully-activated dimer is robust to stochastic perturbations: erroneous deactivation of one monomer is rapidly corrected by the enzyme.

This proposal (and a closely related one by Lisman \cite{lisman1985mechanism}) has been influential in the study of the molecular basis of long-term memory in neural and cellular systems, with particular complexes such as CaMKII being argued to act in this way \cite{stratton2022start}. However, it is not yet known if it has broader implications for the design of \textit{de novo}, synthetic complexes capable of memory and computation.

Besides concrete applications to memory and computation, there is an increasing interest in how nonequilibrium processes, where the kinetics matter more than the energetics, enhance the capability of protein complexes to self-assemble \cite{osat2023non,benoist2025high,PhysRevLett.134.068402}, or to undergo cooperative conformational changes \cite{tu2008nonequilibrium,mahdavi2024flexibility} beyond the equilibrium physics of allostery \cite{monod1965nature,phillips2020molecular}. Cooperativity acts to reduce the number of possible states in which a complex may be found, from a combinatorially large one to a small subset of attractors. In a nonequilibrium context, a protein complex may cycle through states in a way that breaks time reversibility, as in the case of molecular clocks \cite{rust2007ordered}. Topological protection driven by nonequilibrium futile cycles has been recently proposed as a generic mechanism by which protein complexes can undergo dimensionality reduction \cite{murugan2017topologically,PhysRevX.11.031015,agudocanalejo2024topologicalphasesdiscretestochastic}, but whether other nonequilibrium mechanisms can generically achieve this remains an open question.

\begin{figure*}
	\centering
	\includegraphics[scale=1]{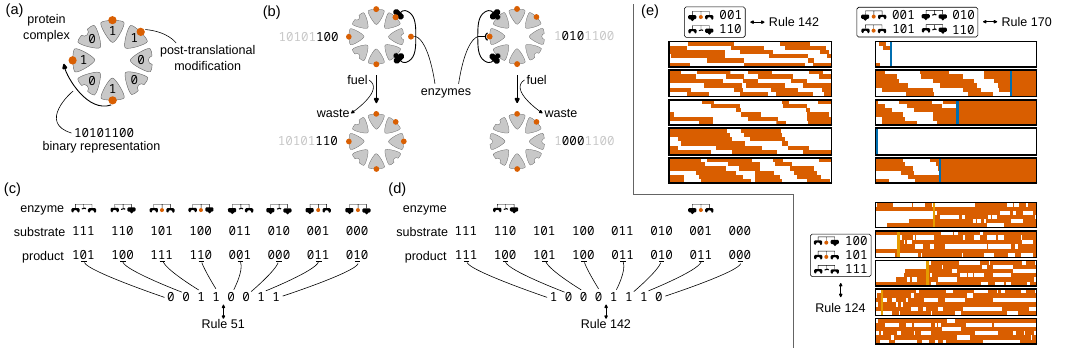}
	\caption{\textbf{Nonequilibrium transitions in a protein complex realise cellular automata rules.}
		(a) Schematic of a protein complex made of identical monomers, each of which can be post-translationally modified (e.g., phosphorylated). We represent the state of the complex by a binary string, which indicates the modification state of each monomer.
		(b) Transitions are catalysed by enzymes that add or remove modifications; enzyme recruitment to a site depends on the modification status of its neighbours.
		(c) The eight possible enzymes can each be represented by a  binary triplet, denoting the state of the monomer triplet to which they preferentially bind (substrate) before the action of the enzyme. The set of enzymes present in the system, and therefore the `rule' (set of allowed transitions), is captured in the Wolfram code which assigns rules a number between 0 and 255. This example with all eight enzymes present corresponds to rule 51.
		(d) Another example, with only two enzymes present, corresponding to rule 142.
		(e) Simulated stochastic trajectories of a protein complex of size $N = 6$ under different rules, showing a nonequilibrium travelling wave (rule 142), localization to one of two single state attractors (rule 170), and equilibrium-like switching between a subset of states (rule 124).
	}\label{fig:1}
\end{figure*}

Here, we study a class of nonequilibrium models that generalise Crick's proposal for molecular memory to complexes of arbitrary size.
We construct a thermodynamically-consistent model of a nonequilibrium protein complex in which each subunit is post-translationally modified with a rate that depends on the state of its two neighbours. In the limit of strong nonequilibrium driving, we map the system to a stochastic variant of cellular automata -- a set of dynamics we term `molecular automata'. By systematically classifying all possible implementations, we highlight features that can be used to store and process information at the nanoscale -- including error-tolerant, multistable attractors that can show nonequilibrium dynamics, long transient timescales, and a programmable sequence of transitions reminiscent of finite-state machines. We discuss how these might be used to construct adaptive, computing devices out of genetically-encoded elements in a living cell.

\section{Model}

\subsection{Context-sensitive enzymes}

We consider a complex made of $N$ identical monomers arranged in a circular complex [Fig.~\ref{fig:1}(a)]. Each monomer can take one of two conformations (0 and 1, representing, e.g., structural conformation or post-translational modifications such as phosphorylation), with the global state of the complex denoted by a binary string of length $N$. The complex transitions between its $2^N$ global states when individual monomers switch their conformation.

As each monomer makes physical contact with two neighbours, its switching rate may in principle depend on the states of its neighbours. We denote a triplet of contiguous monomers in states $i,k,j \in \{0,1\}$ as $ikj$. In an equilibrium setting, the rate $k^{\text{spo}}_{ikj \to i \bar{k}j}$ for a monomer in state $k$ to spontaneously switch into state $\bar{k}$ (the bar denoting logical negation) while surrounded by monomers in states $i$ and $j$ is constrained to satisfy detailed balance,
\begin{equation}
	\frac{k^{\text{spo}}_{ikj \to i \bar{k}j}}{k^{\text{spo}}_{i\bar{k}j \to ikj}} = e^{\beta\left(\epsilon_{ikj} - \epsilon_{i\bar{k}j}\right)},
	\label{eq:detbal}
\end{equation}
where $\beta \equiv (k_B T)^{-1}$ with $k_B$ Boltzmann's constant and $T$ the temperature, and $\epsilon_{ikj}$ is the energy of the three monomers in states $i$, $k$, and $j$ in contact with each other, which may include contributions favouring or penalizing the $ik$ and $kj$ interfaces. Similar models have been previously studied in the context of equilibrium allosteric complexes such as haemoglobin \cite{monod1965nature}. By energetically penalizing interfaces between unequal conformations, such equilibrium models can for example describe allosteric complexes where all monomers prefer to be in the same conformation \cite{wodak2019allostery}.

However, going beyond this equilibrium paradigm, transitions may also be driven out of equilibrium by coupling to a (free) energy source or `fuel', such as a clamped reservoir of cytosolic ATP.
Such a transition arises naturally when it is enabled by an ATP-consuming enzyme (e.g., a kinase), or when the complex itself has ATP-hydrolytic activity (as in AAA-ATPases such as the molecular clock KaiC \cite{tomita2005no}).

In particular, we consider that each change of state is catalysed by a dedicated enzyme $(ikj)$ (note the use of round brackets to distinguish it from the corresponding monomer triplet) which binds to a triplet of monomers and catalyzes the reaction $(ikj) + ikj + \text{fuel} \rightleftarrows (ikj) + i\bar{k}j + \text{waste}$, see Fig.~\ref{fig:1}(b). For example, in this notation a kinase that phosphorylates a monomer only when both neighbours are phosphorylated would be denoted as (101). Local detailed balance in this case demands that the forward and backward rates of this enzymatic reaction satisfy
\begin{equation}
	\frac{k^{{(ikj)}}_{k \to \bar{k}}}{k^{{(ikj)}}_{\bar{k} \to k}} = e^{\beta\left(\epsilon_{ikj} - \epsilon_{i\bar{k}j}+\Delta \mu_{(ikj)}\right)},
	\label{eq:detbal2}
\end{equation}
where $\Delta \mu_{(ikj)}>0$ is the chemical potential difference of the fuel and waste reservoirs to which enzyme $(ikj)$ couples.

In order to focus on the nonequilibrium behaviour enabled by enzymes, we will consider in the following that there is no energetic preference for any of the states of the complex, i.e.~$\epsilon_{ikj}$ is the same for all possible monomer triplets and can be set to zero without loss of generality. Alternatively, this could describe the case where the energetics of the complex are negligible relative to the nonequilibrium driving forces, i.e.~$|\epsilon_{ikj} - \epsilon_{i\bar{k}j}| \ll |\Delta \mu_{(ikj)}|$. Moreover, for simplicity, we consider that all enzymatic reactions couple to the same fuel and waste reservoirs, so that we can set $\Delta \mu_{(ikj)} = \Delta \mu > 0$ for all enzymes.
The rate at which enzyme $(ikj)$ catalyses the forward reaction $k \to \bar{k}$ can then be explicitly written as
\begin{equation}
	k^{(ikj)}_{k \to \bar{k}} = k_{(ikj)} \frac{e^{\beta \Delta \mu}}{e^{\beta \Delta \mu} + 1},
	\label{eq:forwardrate}
\end{equation}
where $k_{(ikj)}$ is the overall kinetic rate of the reaction, which includes the cellular concentration of enzyme $(ikj)$. The rate of the backward reaction $\bar{k} \to k$ is in turn
\begin{equation}
	k^{(ikj)}_{\bar{k} \to k} = k_{(ikj)} \frac{1}{e^{\beta \Delta \mu} + 1},
	\label{eq:backwardrate}
\end{equation}
where we note that the local detailed balance condition (\ref{eq:detbal2}) is automatically satisfied. The denominators in (\ref{eq:forwardrate}) and (\ref{eq:backwardrate}) are chosen to ensure that the reaction rates have an upper bound of $k_{(ikj)}$  even if the roles of fuel and waste were to be reversed and thus $\Delta \mu < 0$.
Note that, importantly, while enzyme $(ikj)$ preferentially drives the reaction $k \to \bar{k}$, another enzyme $(i\bar{k}j)$ can couple to the energy reservoir to drive the reverse reaction $\bar{k} \to k$ (as in the famous `push-pull' regulatory motifs \cite{stadtman1977superiority,detwiler2000engineering}).

\subsection{Strongly-driven limit and molecular automata}

In practice, biological systems are strongly driven ($\beta\Delta \mu \gg 1$), such that reverse reactions do not occur on physiologically-relevant timescales. We therefore take the strongly-driven limit ($\beta \Delta \mu  \to \infty$) and, for simplicity, suppose that each $k_{(ikj)}$ either takes on a constant non-zero value $k_\mathrm{cat}$ or is $0$ (i.e., the enzyme is present or absent). Later in the paper, we will relax these assumptions.

In this limit, the dynamics of a complex is fully specified by the presence or absence of each enzyme type $(ikj)$ (with $k_\text{cat}$ only setting the overall timescale), where we remind the reader that $i,k,j \in \{0,1\}$. As such there are $2^3=8$ possible enzymes, shown schematically in Fig.~\ref{fig:1}(c). Depending on which of these eight enzymes are present or absent, there are $2^8=256$ possible enzyme subsets, each defining a rule (set of allowed transitions for a monomer given its state and that of its two neighbours) that governs the stochastic dynamics of the complex.

This description enables us to draw a parallel with the theory of elementary cellular automata, where each rule can be specified by a Wolfram code \cite{RevModPhys.55.601}: an eight-digit binary number, thus running from 0 to 255, that encodes the fate of the central monomer $k$ in each triplet configuration $ikj$. For example, the rule where all eight enzymes are present is depicted in Fig.~\ref{fig:1}(c) and corresponds to Wolfram code 51. On the other hand, the rule where only $(001)$ and $(110)$ are present is depicted in Fig.~\ref{fig:1}(d) and corresponds to Wolfram code 142. However, unlike well-studied deterministic cellular automata, our system is stochastic: transitions occur probabilistically and asynchronously, with only one monomer switching at a time. Asynchronous cellular automata have been previously studied and their dynamics have been shown to be distinct to those of synchronous ones, and sensitive to the precise scheme by which cells are randomly updated. \review{The strongly-driven limit of our model is analogous to asynchronous cellular automata with random, independent updates, whose dynamics have been previously studied in the large $N$ limit \cite{cornforth2005ordered}, but in the following we focus on small $N$ properties, as well as on deviations from the strongly-driven limit, both relevant to molecular assemblies.}
Because it is particularly well-suited to the description of molecular-scale stochastic processes, we term this model `molecular automata'.

Just as with synchronous automata, however, not all of the 256 rules are dynamically distinct. Rules are subject to three possible symmetry operations: bit-flipping (swapping all zeroes and ones), left-right reflection (interchanging the roles of left and right neighbours), and the combined application of both. Considering only the rules that are distinct from each other up to these three operations, the $256$ rules reduce to $88$ dynamically-distinct rules \cite{li1990structure}.

Famously, despite the simplicity of their local rules, the global dynamics of a cellular automaton can be surprisingly rich. We sought to see if this is true of our molecular analogue. In Fig.~\ref{fig:1}(e) we show representative dynamics as space-time diagrams for a complex with $N=6$ monomers, where the vertical direction represents the state of the complex at a given time, and time progresses along the horizontal direction (see Appendix~\ref{sec:appendix} for details on the simulation method). We observe behaviours as varied as localization into a single state (rule 170), time-reversible equilibrium-like dynamics (rule 124), and time-irreversible nonequilibrium dynamics such as travelling waves (rule 142).

\begin{figure*}
	\centering
	\includegraphics[scale=1]{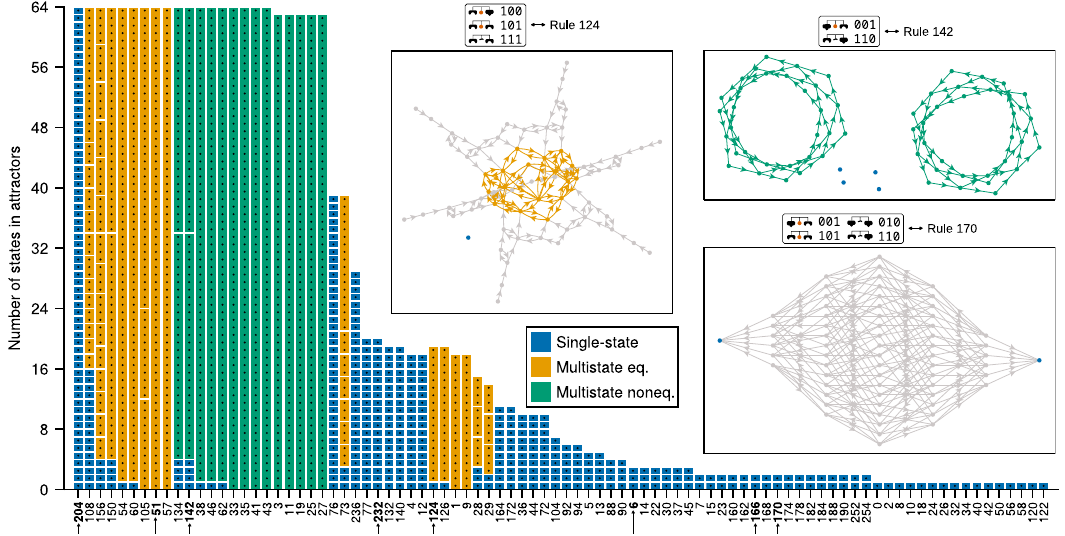}
	\caption{\textbf{Conformations are funnelled into distinct stable attractors.}
		Shown are the number of states in attractors for each of the 88 non-redundant rule-sets (labelled by their Wolfram codes), for a complex of size $N=6$. States in the same attractor are grouped together into boxes, which are coloured to denote the type of attractor. \review{Rules explicitly discussed in Figs.~\ref{fig:1}--\ref{fig:4} have been highlighted.} The insets show three rule-sets as directed graphs, where nodes denote a unique state of the complex, and transitions are catalysed reactions between states. States not in an attractor are in gray, others are coloured. 
	}\label{fig:2}
\end{figure*}

\section{Results}

\subsection{Rule-dependent attractors partition the long-time dynamics of molecular automata}

To understand the long-term behaviour of each rule, we systematically classified the attractors of the associated stochastic dynamics. Each rule defines a Markov process over the $2^N$ global states of the complex, and its attractors span the null-space of the transition rate matrix (see Appendix~\ref{sec:appendix} for details on the computation of attractors).

Attractors are of two broad types: single-state (i.e.~a state that, once reached, cannot be left) and multi-state. We distinguish between two types of the latter: `equilibrium' attractors in which every transition is reversible, and nonequilibrium attractors in which at least one transition is irreversible (in the limit of strong chemical driving which we continue to take in the following unless specified). By definition, attractors are strongly connected, so nonequilibrium attractors must have a non-zero steady state current. Note also that what we call equilibrium attractors are still nonequilibrium in the thermodynamic sense, as the enzymes are coupled to free energy reservoirs. However, in these attractors, for every transition in one direction catalyzed by one enzyme there is also the reverse transition catalyzed by a different enzyme, making all transitions reversible and the dynamics equilibrium-like.

The number of attractors and their nature, as well as the number of states belonging to the attractors, depends on the rule (i.e.~the set of enzymes available) and also on the size of the complex $N$. In Fig.~\ref{fig:2}, we show a rank-ordered plot of the number of states in attractors for all 88 dynamically-distinct rules, for a hexameric complex ($N=6$). Two extreme cases correspond to two `trivial' rules: when no enzymes are present (rule 204), no transitions are allowed, and there are therefore $2^6=64$ trivial single-state attractors; when all eight enzymes are present (rule 51), all transitions are allowed, and there is a single equilibrium attractor containing all 64 states.

For other rules, the behaviour can be highly non-trivial. Three examples are shown in the inset of Fig.~\ref{fig:2}. For rule 142, we find four single-state attractors, and two multistate nonequilibrium attractors containing all remaining 60 states of the complex, corresponding to two different types of travelling waves of conformational changes along the length of the complex (see Fig.~\ref{fig:1}(e) for simulated stochastic trajectories within the nonequilibrium attractors). For rule 170, on the other hand, we find only two single-state attractors,corresponding to the complex being in the $\mathbf{0}$ (all zeroes) or $\mathbf{1}$ (all ones) state. The rule `funnels' any other state into these two attractors (see Fig.~\ref{fig:1}(e) for stochastic simulations). Thus, molecular automata can effectively reduce the size of the configuration space, providing an example of purely nonequilibrium allostery (cooperativity). How much any given rule reduces the configuration space can be quantified by the total number of states present in the attractors (total height of bars in Fig.~\ref{fig:2}). Lastly, rule 124 shows two attractors, a single-state one and a multistate equilibrium one containing only a fraction of the remaining states. Examples of the dynamics within the equilibrium multistate attractor can be seen in Fig.~\ref{fig:1}(e).

\begin{figure}
	\centering
	\includegraphics[scale=1]{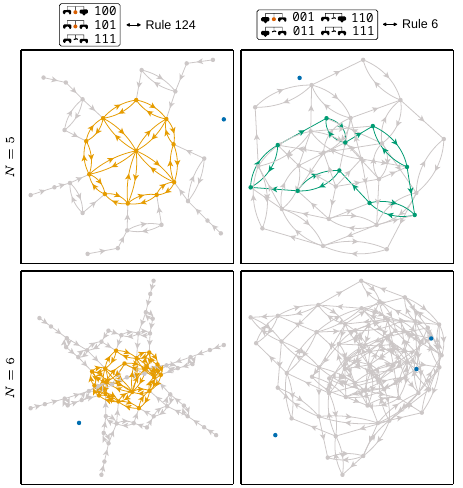}
	\caption{\textbf{Some rules can be highly sensitive to the size of the complex, while others remain robust.}
		Transition graphs for rules 124 and 6, for complex sizes $N=5$ and $N=6$. Colours are as in Fig.~\ref{fig:2}. The attractor structure of rule 124 is robust, showing an equilibrium multistate attractor and a single-state attractor in both cases. Rule 6 is highly sensitive to whether the complex size is odd or even, showing a nonequilibrium cycle coexisting with a single-state attractor for $N=5$, and just three single-state attractors for $N=6$.
	}\label{fig:3plus}
\end{figure}

As mentioned above, the behaviour of a rule depends also on the size $N$ of the molecular complex. In particular, we find that there are strong parity effects, with complexes with even and odd numbers of monomers showing markedly different behaviours. For instance, a hexamer evolving under rule 6 tends towards one of two single-state attractors. In contrast, a pentamer evolving under the same rule is driven towards a single nonequilibrium cyclic attractor (Fig.~\ref{fig:3plus}, right). This is not always the case, however, as e.g.~both hexamers and pentamers under rule 124 evolve towards an equilibrium multistate attractor (Fig.~\ref{fig:3plus}, left). The rank-ordered plot for the number of states present in the attractors for $N=5$ is shown in SM Fig.~S1 \cite{suppmat}. {Notably, Rule 142 encodes non-equilibrium attractors at all $N$ (SM Fig.~S4 \cite{suppmat}); given its relative simplicity (only two enzymes), it may be useful in constructing synthetic analogues of oscillatory complexes such as KaiC.}

\subsection{Relaxation kinetics and molecular stopwatches}

\begin{figure*}
	\centering
	\includegraphics[scale=1]{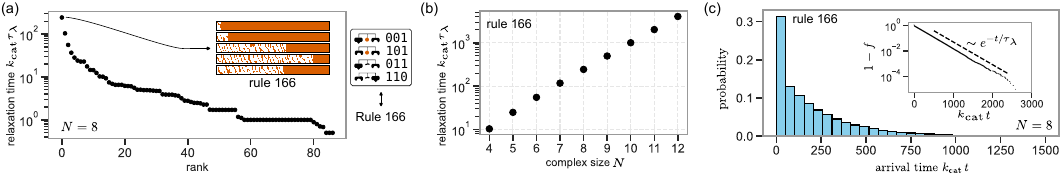}
	\caption{\textbf{Relaxation kinetics and molecular stopwatches}  (a) Rank-ordered plot of inverse spectral gaps, i.e.~longest relaxation times $\tau_\lambda$, across all 88 rules for an octamer with $N=8$. Inset: five replicas of stochastic simulations for rule 166, which has the longest relaxation time. (b) Relaxation time as a function of complex size $N$ for rule 166, showing exponential scaling. (c) Distribution of arrival times to the attractor $\mathbf{1}$ for a population of complexes with $N=8$ initialized in a random state, for rule 166. Inset: the fraction of complexes that have not reached the attractor scales as $e^{-t/\tau_\lambda}$.
	}\label{fig:3}
\end{figure*}

Rules with a small set of single-state attractors (e.g.~one or two) require the system to funnel down a large configuration space before reaching the attractor. The typical timescale required to reach the attractor can be much larger than the timescale $k_\mathrm{cat}^{-1}$ of individual transitions.

To quantify these relaxation times, we calculate the spectral gap $\tau_\lambda^{-1} \equiv \lvert \text{Re}( \lambda ) \rvert$, where $\lambda$ is the eigenvalue of the system's transition matrix with real part closest to (but not equal to) zero. Note that, for our continuous time Markov process, the transition matrix will have $2^N$ eigenvalues, which may be either zero (corresponding to the steady state attractors of the system) or have negative real parts, corresponding to relaxation modes. The spectral gap therefore gives the inverse of the slowest relaxation timescale $\tau_\lambda$ in the system.

A rank-ordered plot of the relaxation times $\tau_\lambda$ for all rules is shown in Fig.~\ref{fig:3}(a), for an octamer with $N=8$. We find a wide spread, with values as large as $\tau_\lambda \simeq 244$ (in units of $k_\mathrm{cat}^{-1}$) for rule 166. Interestingly, we find that the relaxation times for this rule increase exponentially with the complex size $N$, as shown in Fig.~\ref{fig:3}(b). \review{Of the rules with the longest relaxation times in Fig.~\ref{fig:3}(a), this appears to be a unique feature of Rule 166 (see SM Fig.~S5 \cite{suppmat}).}

Such long relaxation times could be exploited to build a `molecular stopwatch', using populations of complexes and employing the fraction of complexes $f$ that have reached the attractor as a timekeeper. A numerical experiment demonstrating this for $N=8$ is shown in  Fig.~\ref{fig:3}(c), which shows the histogram of arrival times to the attractor $\mathbf{1}$ (all ones) for a population of complexes starting in random initial states. At long times, the fraction $1-f$ of complexes that have not yet reached the attractor scales as $1-f \sim \exp(-t/\tau_\lambda)$ (see inset). \review{As $\tau_\lambda$ can be much larger than the molecular timescale $1/k_{\text{cat}}$, such a stopwatch could in principle be built for a range of timescales simply by varying $N$.}

\subsection{Error correction and stochastic switching}

Rules with more than one attractor can be used to store memory. In particular, there are four rules that have the $\mathbf{0}$ (all zeroes) and $\mathbf{1}$ (all ones) states as attractors, at least one incoming transition into each of them, and are moreover invariant to the bit-flipping operation (rules 170, 178, and 232) or invariant to the combined bit-flipping and mirror symmetry operations (rule 184). These symmetric rules are particularly useful to implement a bistable memory, where the $\mathbf{0}$ and the $\mathbf{1}$ states are attractors and the dynamics are symmetric with respect to how the system can reach these states.

An interesting question is then to consider how good these different rules are at maintaining this memory, and in particular how capable they are of error-correcting undesirable bit-flips. Indeed, error correction becomes essential if the nonequilibrium driving force is not infinite or if non-enzymatic, spontaneous transitions are allowed, as in both cases it becomes possible for the system to jump out of an attractor. We will consider both of these possibilities below.

Before doing so, however, we note it is possible to gauge how good the different rules are at keeping memory while remaining within the strongly-driven, enzyme-driven limit. To this end we quantify  the splitting probabilities of any given starting state, i.e.~the probability that given a starting state the system ends in the $\mathbf{0}$, $\mathbf{1}$, or in a different attractor (see Appendix~\ref{sec:appendix} for details on the computation of splitting probabilities). These are shown in SM Fig.~S2 \cite{suppmat}. One should demand from a good error-correcting rule to have a high splitting probability towards the $\mathbf{0}$ or $\mathbf{1}$ state in response to small numbers of bit flips away from the corresponding state.

\begin{figure*}
    \centering
    \includegraphics[scale=1]{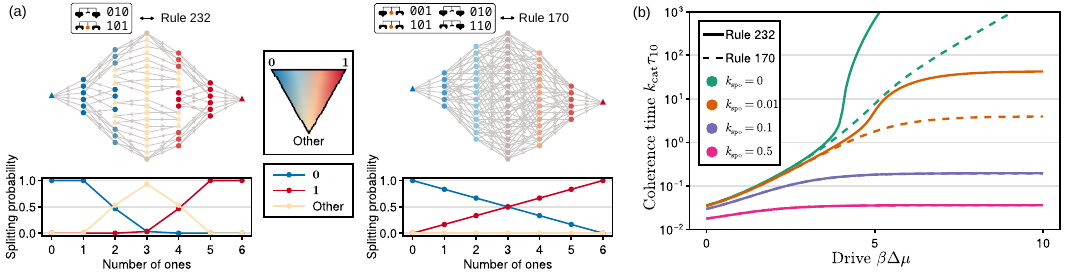}
    \caption{\textbf{Error correction under finite driving and spontaneous reactions.} (a) Splitting probability towards the $\textbf{0}$ and $\textbf{1}$ attractors, as well as towards any other attractor, for rules 232 and 170, as a function of the number of ones in the complex. Rule 232 demonstrates perfect error correction under single bit flip errors, with a splitting probability of 1 away from $\textbf{0}$ or $\textbf{1}$. (b) Coherence time $\tau_{10}$, representing how long it takes $10\%$ of complexes to exit the $\textbf{0}$ state, for both rules as a function of the nonequilibrium driving $\Delta \mu$, for several values of the spontaneous transition rate $k_\mathrm{spo}$ (in units of $k_\mathrm{cat}$). 
    }\label{fig:4}
\end{figure*}

\review{For illustration purposes, we focus here on rule 232 and rule 170, which have promising error-correcting capabilities,} see Fig.~\ref{fig:4}(a). Rule 232 has perfect error correction (splitting probability of 1) under single bit flips, and is therefore a generalization of Crick's hypothetical error-correcting mechanism (only defined for $N=2$ and only capable of correcting the $\mathbf{1}$ state \cite{crick1984neurobiology}) to error-correction of both the $\mathbf{0}$ and the $\mathbf{1}$ states for arbitrary $N$. However, this rule becomes fragile at large numbers of bit flips, where there is a high splitting probability towards spurious attractors that are neither $\mathbf{0}$ nor $\mathbf{1}$. On the other hand, rule 170 has no spurious attractors and has a linearly decreasing splitting probability with the number of bit flips, but lacks perfect error correction under single bit flips. \review{We find that these results hold even when the complex size $N$ is changed (see SM Fig.~S6 \cite{suppmat}).}

We now proceed to test the error-correcting capabilities of these two rules by allowing for a finite driving force $\Delta \mu$ \review{(thus allowing a reverse transition for every catalysed transition)}, and a non-zero rate for the non-enzymatic spontaneous transitions $k^{\text{spo}}_{ikj \to i \bar{k}j}=k^{\text{spo}}_{i\bar{k}j \to ikj}=k_\text{spo}$ \review{(thus allowing any monomer to change state at rate $k_\text{spo}$)}. In Fig.~\ref{fig:4}(b), we quantify the time $\tau_{10}$ it takes for $10 \%$ of a population of complexes initialized in state $\mathbf{0}$ to transition out of it, or coherence time, as a function of the driving force $\Delta \mu$, for several values of the spontaneous transition rate $k_\text{spo}$. In the absence of spontaneous transitions ($k_\text{spo}=0$), for both rules the coherence time increases exponentially with increasing $\Delta \mu$, but rule 232 (perfect single bit flip correction) performs significantly better than rule 170, leading to much larger timescales. For finite but small $k_\text{spo}$, the coherence time saturates to a finite value with increasing $\Delta \mu$, but still rule 232 significantly outperforms rule 170. When $k_\text{spo}$ approaches the same order of magnitude as the enzymatic rate $k_\mathrm{cat}$, however, the error-correction capabilities sharply drop and both rules perform equally poorly. \review{For completeness, we show equivalent plots for the other two error-correcting rules, 184 and 178, in SM Fig.~S3 \cite{suppmat}. Their performance is comparable to that of rule 170, i.e.~significantly worse than that of rule 232.}

Lastly, we note that, under finite driving or in the presence of spontaneous transitions, multistability becomes a rather generic feature of molecular automata, beyond the bistable systems just discussed. \review{In particular, finite driving enables reverse catalyzed transitions, and thus allows for the stochastic switching between attractors as long as there is a path connecting them in the undirected version of the transition graph, as for the two attractors in rule 170 (Fig.~\ref{fig:2}). Spontaneous transitions allow for stochastic switching even between attractors that are disconnected in the undirected version of the transition graph, as in the case of rules 142 and 124 (Fig.~\ref{fig:2}), because any monomer can now switch state and thus all states of the complex become connected. Note also that} stochastic switching can occur between attractors of a different nature. For example, in the presence of spontaneous transitions, rule 142 displays spontaneous switching between static phases (single-state attractor) and travelling waves (nonequilibrium attractor).

\subsection{Protein complexes as finite-state machines}

Up until now, we have focused on understanding the dynamics of protein complexes under the application of any given rule. Among other features, we have found that under certain rules and for certain starting conditions we can, in the strongly-driven, purely-enzymatic limit, guarantee that the system will deterministically end up at a particular single-state attractor.

This suggests a strategy for precisely controlling the state of the complex at any given time by judiciously applying a sequence of rules. Indeed, for any given global state of the complex, we can determine all rules that, starting from that state, will deterministically lead to a new single-state attractor (see Appendix~\ref{sec:appendix} for details).  In this way, the protein complex becomes a deterministic `finite-state machine', which can transition from one state to another in response to some input that leads to a change in the set of enzymes present (via, e.g., inducible promoters in an \textit{in vivo} context). Finite-state machines correspond to one of the most basic models of computation, capable of implementing combinational logic but not as general as a Turing machine \cite{hopcroft2001introduction}.

\begin{figure*}
    \centering
    \includegraphics[scale=1]{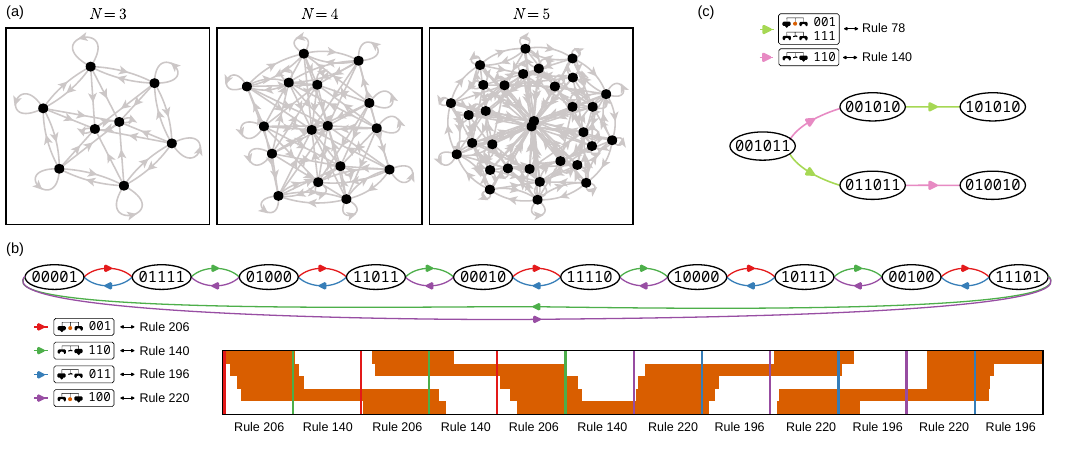}
    \caption{\textbf{Building finite-state machines by inducible enzyme sets.} (a) Full finite-state transition diagrams for $N=3$, $N=4$, and $N=5$. Each arrow corresponds to application of a specific rule, i.e.~to the presence of a given enzyme set, and each node to a specific state of the complex. (b) Top: Detailed view of a `counter' module in the $N=5$ finite-state machine. Alternating application of a pair of rules (206 and 140) deterministically changes the state of the complex. Alternating application of second pair of rules (196 and 220) can reverse the state of the counter. Bottom: Stochastic simulation showing the counter in action. (c) Detailed view of an `order-recorder' module in the $N=6$ finite-state machine. Starting from a given state, the system can tell in which temporal order two inputs (rules 78 and 140) have appeared, as they take the system to distinct states. 
    }\label{fig:5}
\end{figure*}

We stress that the finite-state machine behaviour is not a property of any given rule, but rather arises from the combination of all of them. As such, the emergent finite-state machine depends only on the size $N$ of the complex. The finite-state machines emerging for $N=3$, $4$, and $5$ are shown in Fig.~\ref{fig:5}(a), and become significantly more complex with increasing $N$. Among these complex diagrams, one can identify particular features, or modules, with computational capabilities. As one example, as seen in Fig.~\ref{fig:5}(b), the finite-state machine for $N=5$ contains a `counter' module, with the state of the complex depending on how many times the system has switched between rules 206 and 140, until it resets after 10 events. 
As another example, as seen in Fig.~\ref{fig:5}(c), the finite-state machine for $N=6$ contains an `order-recorder' module, which can tell apart the temporal order in which two different inputs (corresponding to rules 78 and 140\review{, which could be coupled to e.g.~two distinct environmental stimuli}) have been presented to the protein complex. 

\review{Delineating the full spectrum of computational behaviour that a particular molecular automaton can exhibit requires exhaustive enumeration of the finite-state machine graph, which can be computationally challenging for large $N$. However, realising particular computational tasks can sometimes be designed \textit{a priori}. For instance, the counter realised by rules 206, 140, 196, and 220 can be implemented in any molecular automaton of $N>2$ when initialised in a state with a single flipped bit (i.e. a single monomer in state 1 or 0). The counter will be able to count $N$ stimuli when $N$ is even, and $2N$ stimuli when $N$ is odd (see SM Fig.~S7 \cite{suppmat}). The order-recorder behaviour, however, is more complex, and will only occur via rules 78 and 140 for molecular automata with $N \geq 6$ (see SM Fig.~S8 \cite{suppmat}).}

From an engineering perspective, this behaviour enables precise control over the global state of a protein complex, or an entire population of them. From a synthetic biology perspective, it opens up the possibility for programming cellular responses that do not depend only on the present stimulus, but on a whole sequence of past stimuli, endowing them with a protein-scale memory (see \cite{roquet2016synthetic} for a DNA-based variant). Future work may explore the kinds of logical computations that can be implemented within this paradigm, or extensions of it.

\section{Discussion}

We have analysed a generic model of a multicomponent protein assembly, subject to driven, enzymatic reactions that switch the state of each monomer in a context-dependent manner. In the limit of strong chemical driving, far from detailed balance, we find that these systems naturally implement a stochastic, asynchronous variant of cellular automata rules. These molecular automata can have multiple stable attractors, which exhibit either time-reversible equilibrium-like dynamics or time-irreversible nonequilibrium dynamics such as travelling waves. We find long transients that can be used as molecular stopwatches, and determine which rule-sets are capable of error-tolerant memory. Finally, by dynamically switching rule-sets (as implementable by, e.g., genetically inducible enzymes), molecular automata could be pre-wired to realise a finite-state machine -- suggesting ways to build computable elements out of proteins in living cells.

Our work opens up several avenues for generalisation. We have studied the case in which each monomer has only a single modification site -- in contrast, natural proteins often have, e.g., several independently phosphorylatable sites. Increasing the number of sites could lead to a new class of dynamics, and more expressive computational abilities. Further, while we have indicated how finite-state machines might be implemented, it would be particularly interesting to construct concrete realisations, and to determine which finite-state machines can (and cannot) be realised in this system. Finally, in our work, we have supposed that the dynamics of each macromolecular complex is independent of all others -- but one could easily imagine situations in which complexes interact via, e.g., the activation of enzymes, or monomer exchange \cite{ito2007autonomous}. Such feedback should lead to a new class of dynamics not realisable in the current setting.

While we have taken inspiration from naturally-occuring biological systems, it is unclear if any extant biomolecular complex implements a molecular automaton. Many of the ingredients are, however, naturally present. There are numerous examples of homomeric protein complexes with modifiable monomers; for instance, the AAA-ATPase family that includes the hexameric KaiC molecular clock and the pentameric DNA clamp loader. In other contexts, post-translationally modifying enzymes such as kinases often act in a context-dependent manner (e.g., recruited to a substrate only when some other site is phosphorylated, or in a particular structural conformation). \review{If biology were to realize and make use of molecular automata, they might be found as a non-genetic memory (error-correcting rules), as an advanced sensing mechanism (counter and order-recorder modules) or, for more complex finite-state machine modules, as an information processing and decision-making apparatus that integrates many signals to regulate the synthesis and function of other protein complexes.}

\review{In any case, even if biology does not exploit these components in the way we have suggested, molecular automata could be built via the current toolkit of synthetic biology (i.e.~re-wiring existing components).}
Indeed, it is not inconceivable that such systems can now be designed from scratch. Recent advances in protein design have lead not only to the realisation of \textit{de novo} folds \cite{kuhlman2003design}, but also to assemble individual protein subunits into complexes with a specified stoichiometry and arrangement \cite{king2014accurate}, and with the capability to switch conformation in response to ligand binding \cite{pillai2024novo}. It is only a short step to functionalise these complexes by, e.g., including post-translationally modifiable sites that couple to subunit conformation and potentially initiate downstream signalling. These advances suggest that building synthetic, functional systems at the molecular scale out of designed protein components is within reach; our results could therefore be used as design principles to target a desired behaviour.

\review{Outside of a biological context, our study is relevant to the wider field of cellular automata. In its simplest avatar (strongly driven, fixed rule-set), our model recapitulates that of asynchronous cellular automata with random, independent updates \cite{cornforth2005ordered}. While these are often studied dynamically in the large-$N$ limit, we present instead a global enumeration and classification of attractors for small $N$ (Fig.~2). Determining the attractor structure at larger $N$ remains an interesting problem for future work to consider.} 

\review{Moreover, our systematic derivation from an underlying, thermodynamically-consistent model brings with it new features: for instance, distinguishing between the two types of `erroneous' transitions (i.e., arising from the finite non-equilibrium driving of the reaction, and from spontaneous chemical transitions) as in Fig.~5(b). In addition, the underlying biological inspiration motivates studying novel aspects of cellular automata, such as their behaviour when the rules themselves change dynamically, leading to finite-state machines.}

Finally, our results can be seen as generalising classic models of allostery to intrinsically out-of-equilibrium dynamics. The classic Monod-Wyman-Changeux model of allostery \cite{monod1965nature} has been widely influential and applied to situations far from its initial remit \cite{phillips2020molecular}. There, individual components of a larger complex interact energetically to give rise to behaviour not possible in a single monomer. Our model is similar, yet the effect of neighbouring monomers in a complex is to modulate the kinetics rather than the energetics of some chemical landscape -- giving rise to the behaviour we have analysed. In this, our study is part of a larger, recent realisation that out-of-equilibrium dynamics at the molecular scale can give rise to qualitatively new information-processing capabilities \cite{benoist2025high, mahdavi2024flexibility}. We look forward to more theoretical and eventually experimental realisations of this idea. 

\acknowledgments

J.K. was supported by a doctoral studentship from the Department of Physics and Astronomy at UCL.

\section*{Data Availability}

The data and code that support the findings of this article are openly available \cite{Kocka2026}.

\appendix

\section{Computational details}
\label{sec:appendix}

For a fixed rule-set and complex size $N$, the stochastic dynamics of the system are specified by the Master equation:
\begin{equation}\label{eq:master}
    \partial_t p_i = \sum_j (R_{ij} p_j - R_{ji} p_i) = \sum_j W_{ij} p_j
\end{equation}
\noindent where $R_{ij}$ is the transition rate from state $j$ to state $i$ and $W_{ij} = R_{ij} - \delta_{ij} \sum_k R_{ki}$ is the transition matrix. For each system, we explicitly construct the transition matrix by enumerating all allowed transitions between each pair of the $2^N$ states $i$ and $j$. All rates are expressed in units of $k_{\mathrm{cat}}$.
\review{Constructing this matrix is the most computationally intensive step of our analysis, as the necessary time scales with $N2^N$.
For example, generating these matrices for all 88 distinct rules takes ${\sim}0.04~\si{s}$ at $N=6$ but ${\sim}3~\si{min}$ at $N=15$.}

The long-time attractors of the dynamics are, in principle, accessible by analysing the null-space of the transition matrix. However, we instead opt for a simpler graph theoretic approach, implemented using the Graphs.jl library in Julia. We first use Tarjan's algorithm to find all strongly-connected components -- subgraphs in which every state is accessible from any other. The attractors of the system are those strongly-connected components that have no outgoing transitions -- which we identify by direct inspection.
\review{In terms of performance, finding the attractors scales in the worst case as $N2^N$, though we generally see much better scaling when analysing all 88 distinct rules.
Classifying the attractors as either single-state or multistate equilibrium-like or nonequilibrium is done through direct inspection of the edges, and scales quadratically with the number of edges within each attractor.}

To simulate stochastic trajectories of an individual protein complex, \review{for the} space-time diagrams in the Figures, we use the Gillespie algorithm \cite{gillespie2007stochastic}.
\review{This is implemented using a precomputed list of neighbours, and thus the time taken for every Gillespie step only scales linearly with the number of neighbours of the current state, which is at most $N$.
All the Gillespie simulations we performed for the figures in this paper took on the order of milliseconds.}

To calculate the coherence times $\tau_{10}$ in Fig.~\ref{fig:4} we integrate Eq.~(\ref{eq:master}) to obtain
\begin{equation}
    \vb{p}(t) = e^{Wt} \vb{p}(0)
\end{equation}
where we set $\vb{p}(0)$ to correspond to the system being in $\mathbf{0}$ with probability 100\% and then use a numerical solver to find the value of $t$ for which that probability reduces to 90\%.

To compute the splitting probabilities in Fig.~\ref{fig:3}, we construct the matrix $T_{ij}$ from the transition rates:
$$T_{ij} = \frac{R_{ij}}{\sum_{k} R_{kj}}$$
whose value is the probability that the system ends up in state $i$ from state $j$ after a single transition. Raising $T_{ij}$ to a positive integer power generalises this interpretation: $\left[ T^n \right]_{ij}$ is the probability that the system reaches $i$ from $j$ after $n$ transitions. To compute the splitting probability $S_{ai}$ of reaching absorbing state $a$ from some state $i$ we therefore sum over $n$:
$$S_{ai} = \left[ I + T + T^2 + ... \right]_{ai} = \left[ (I - T)^{-1} \right]_{ai}$$
where $I$ is the identity matrix.

To enumerate the finite state machines shown in Fig.~\ref{fig:5}, we modify the procedure for splitting probabilities as follows. First, for fixed $N$, we enumerate the set of all single-state attractors $\mathcal{A}_1$ across all rule-sets. Then, we enumerate all pairs $(a, b) \in \mathcal{A}_1$ such that there exists some rule-set $r$ for which a complex that starts in state $a$ ends up in state $b$ with probability $1$ (i.e. deterministically). To do so efficiently, we use the same procedure to compute the splitting probability from $a$ to $b$ under rule-set $r$, but first replace each multi-state attractor of $r$ with a single-state attractor (to ensure that $I - T$ is invertible).

All identified pairs $(a,b)$ are assembled into a graph, where an edge between states denotes the existence of at least one rule-set that maps $a$ to $b$ deterministically. We manually inspected each of these `finite-state machine' graphs to identify examples of a counter and order-recorder modules.

\vspace{3em}

\bibliography{bibliography}

\begin{thebibliography}{43}%
\makeatletter
\providecommand \@ifxundefined [1]{%
 \@ifx{#1\undefined}
}%
\providecommand \@ifnum [1]{%
 \ifnum #1\expandafter \@firstoftwo
 \else \expandafter \@secondoftwo
 \fi
}%
\providecommand \@ifx [1]{%
 \ifx #1\expandafter \@firstoftwo
 \else \expandafter \@secondoftwo
 \fi
}%
\providecommand \natexlab [1]{#1}%
\providecommand \enquote  [1]{``#1''}%
\providecommand \bibnamefont  [1]{#1}%
\providecommand \bibfnamefont [1]{#1}%
\providecommand \citenamefont [1]{#1}%
\providecommand \href@noop [0]{\@secondoftwo}%
\providecommand \href [0]{\begingroup \@sanitize@url \@href}%
\providecommand \@href[1]{\@@startlink{#1}\@@href}%
\providecommand \@@href[1]{\endgroup#1\@@endlink}%
\providecommand \@sanitize@url [0]{\catcode `\\12\catcode `\$12\catcode
  `\&12\catcode `\#12\catcode `\^12\catcode `\_12\catcode `\%12\relax}%
\providecommand \@@startlink[1]{}%
\providecommand \@@endlink[0]{}%
\providecommand \url  [0]{\begingroup\@sanitize@url \@url }%
\providecommand \@url [1]{\endgroup\@href {#1}{\urlprefix }}%
\providecommand \urlprefix  [0]{URL }%
\providecommand \Eprint [0]{\href }%
\providecommand \doibase [0]{https://doi.org/}%
\providecommand \selectlanguage [0]{\@gobble}%
\providecommand \bibinfo  [0]{\@secondoftwo}%
\providecommand \bibfield  [0]{\@secondoftwo}%
\providecommand \translation [1]{[#1]}%
\providecommand \BibitemOpen [0]{}%
\providecommand \bibitemStop [0]{}%
\providecommand \bibitemNoStop [0]{.\EOS\space}%
\providecommand \EOS [0]{\spacefactor3000\relax}%
\providecommand \BibitemShut  [1]{\csname bibitem#1\endcsname}%
\let\auto@bib@innerbib\@empty
%</preamble>
\bibitem [{\citenamefont {Hopfield}(1982)}]{hopfield1982neural}%
  \BibitemOpen
  \bibfield  {author} {\bibinfo {author} {\bibfnamefont {J.~J.}\ \bibnamefont
  {Hopfield}},\ }\bibfield  {title} {\bibinfo {title} {Neural networks and
  physical systems with emergent collective computational abilities.},\
  }\href@noop {} {\bibfield  {journal} {\bibinfo  {journal} {Proceedings of the
  national academy of sciences}\ }\textbf {\bibinfo {volume} {79}},\ \bibinfo
  {pages} {2554} (\bibinfo {year} {1982})}\BibitemShut {NoStop}%
\bibitem [{\citenamefont {Binder}\ and\ \citenamefont
  {Young}(1986)}]{RevModPhys.58.801}%
  \BibitemOpen
  \bibfield  {author} {\bibinfo {author} {\bibfnamefont {K.}~\bibnamefont
  {Binder}}\ and\ \bibinfo {author} {\bibfnamefont {A.~P.}\ \bibnamefont
  {Young}},\ }\bibfield  {title} {\bibinfo {title} {Spin glasses: Experimental
  facts, theoretical concepts, and open questions},\ }\href
  {https://doi.org/10.1103/RevModPhys.58.801} {\bibfield  {journal} {\bibinfo
  {journal} {Rev. Mod. Phys.}\ }\textbf {\bibinfo {volume} {58}},\ \bibinfo
  {pages} {801} (\bibinfo {year} {1986})}\BibitemShut {NoStop}%
\bibitem [{\citenamefont {Stern}\ \emph {et~al.}(2017)\citenamefont {Stern},
  \citenamefont {Pinson},\ and\ \citenamefont {Murugan}}]{PhysRevX.7.041070}%
  \BibitemOpen
  \bibfield  {author} {\bibinfo {author} {\bibfnamefont {M.}~\bibnamefont
  {Stern}}, \bibinfo {author} {\bibfnamefont {M.~B.}\ \bibnamefont {Pinson}},\
  and\ \bibinfo {author} {\bibfnamefont {A.}~\bibnamefont {Murugan}},\
  }\bibfield  {title} {\bibinfo {title} {The complexity of folding self-folding
  origami},\ }\href {https://doi.org/10.1103/PhysRevX.7.041070} {\bibfield
  {journal} {\bibinfo  {journal} {Phys. Rev. X}\ }\textbf {\bibinfo {volume}
  {7}},\ \bibinfo {pages} {041070} (\bibinfo {year} {2017})}\BibitemShut
  {NoStop}%
\bibitem [{\citenamefont {Murugan}\ \emph {et~al.}(2015)\citenamefont
  {Murugan}, \citenamefont {Zeravcic}, \citenamefont {Brenner},\ and\
  \citenamefont {Leibler}}]{murugan2015multifarious}%
  \BibitemOpen
  \bibfield  {author} {\bibinfo {author} {\bibfnamefont {A.}~\bibnamefont
  {Murugan}}, \bibinfo {author} {\bibfnamefont {Z.}~\bibnamefont {Zeravcic}},
  \bibinfo {author} {\bibfnamefont {M.~P.}\ \bibnamefont {Brenner}},\ and\
  \bibinfo {author} {\bibfnamefont {S.}~\bibnamefont {Leibler}},\ }\bibfield
  {title} {\bibinfo {title} {Multifarious assembly mixtures: Systems allowing
  retrieval of diverse stored structures},\ }\href@noop {} {\bibfield
  {journal} {\bibinfo  {journal} {Proceedings of the National Academy of
  Sciences}\ }\textbf {\bibinfo {volume} {112}},\ \bibinfo {pages} {54}
  (\bibinfo {year} {2015})}\BibitemShut {NoStop}%
\bibitem [{\citenamefont {Khona}\ and\ \citenamefont
  {Fiete}(2022)}]{khona2022attractor}%
  \BibitemOpen
  \bibfield  {author} {\bibinfo {author} {\bibfnamefont {M.}~\bibnamefont
  {Khona}}\ and\ \bibinfo {author} {\bibfnamefont {I.~R.}\ \bibnamefont
  {Fiete}},\ }\bibfield  {title} {\bibinfo {title} {Attractor and integrator
  networks in the brain},\ }\href@noop {} {\bibfield  {journal} {\bibinfo
  {journal} {Nature Reviews Neuroscience}\ }\textbf {\bibinfo {volume} {23}},\
  \bibinfo {pages} {744} (\bibinfo {year} {2022})}\BibitemShut {NoStop}%
\bibitem [{\citenamefont {Waddington}(2014)}]{waddington2014strategy}%
  \BibitemOpen
  \bibfield  {author} {\bibinfo {author} {\bibfnamefont {C.~H.}\ \bibnamefont
  {Waddington}},\ }\href@noop {} {\emph {\bibinfo {title} {The strategy of the
  genes}}}\ (\bibinfo  {publisher} {Routledge},\ \bibinfo {year}
  {2014})\BibitemShut {NoStop}%
\bibitem [{\citenamefont {Raju}\ \emph {et~al.}(2023)\citenamefont {Raju},
  \citenamefont {Xue},\ and\ \citenamefont {Leibler}}]{raju2023theoretical}%
  \BibitemOpen
  \bibfield  {author} {\bibinfo {author} {\bibfnamefont {A.}~\bibnamefont
  {Raju}}, \bibinfo {author} {\bibfnamefont {B.}~\bibnamefont {Xue}},\ and\
  \bibinfo {author} {\bibfnamefont {S.}~\bibnamefont {Leibler}},\ }\bibfield
  {title} {\bibinfo {title} {A theoretical perspective on waddington’s
  genetic assimilation experiments},\ }\href@noop {} {\bibfield  {journal}
  {\bibinfo  {journal} {Proceedings of the National Academy of Sciences}\
  }\textbf {\bibinfo {volume} {120}},\ \bibinfo {pages} {e2309760120} (\bibinfo
  {year} {2023})}\BibitemShut {NoStop}%
\bibitem [{\citenamefont {Ptashne}(1992)}]{ptashne}%
  \BibitemOpen
  \bibfield  {author} {\bibinfo {author} {\bibfnamefont {M.}~\bibnamefont
  {Ptashne}},\ }\href@noop {} {\emph {\bibinfo {title} {A genetic switch}}}\
  (\bibinfo  {publisher} {Blackwell Scientific Publications},\ \bibinfo {year}
  {1992})\BibitemShut {NoStop}%
\bibitem [{\citenamefont {Ozbudak}\ \emph {et~al.}(2004)\citenamefont
  {Ozbudak}, \citenamefont {Thattai}, \citenamefont {Lim}, \citenamefont
  {Shraiman},\ and\ \citenamefont
  {Van~Oudenaarden}}]{ozbudak2004multistability}%
  \BibitemOpen
  \bibfield  {author} {\bibinfo {author} {\bibfnamefont {E.~M.}\ \bibnamefont
  {Ozbudak}}, \bibinfo {author} {\bibfnamefont {M.}~\bibnamefont {Thattai}},
  \bibinfo {author} {\bibfnamefont {H.~N.}\ \bibnamefont {Lim}}, \bibinfo
  {author} {\bibfnamefont {B.~I.}\ \bibnamefont {Shraiman}},\ and\ \bibinfo
  {author} {\bibfnamefont {A.}~\bibnamefont {Van~Oudenaarden}},\ }\bibfield
  {title} {\bibinfo {title} {Multistability in the lactose utilization network
  of escherichia coli},\ }\href@noop {} {\bibfield  {journal} {\bibinfo
  {journal} {Nature}\ }\textbf {\bibinfo {volume} {427}},\ \bibinfo {pages}
  {737} (\bibinfo {year} {2004})}\BibitemShut {NoStop}%
\bibitem [{\citenamefont {Zhu}\ \emph {et~al.}(2022)\citenamefont {Zhu},
  \citenamefont {del Rio-Salgado}, \citenamefont {Garcia-Ojalvo},\ and\
  \citenamefont {Elowitz}}]{zhu2022synthetic}%
  \BibitemOpen
  \bibfield  {author} {\bibinfo {author} {\bibfnamefont {R.}~\bibnamefont
  {Zhu}}, \bibinfo {author} {\bibfnamefont {J.~M.}\ \bibnamefont {del
  Rio-Salgado}}, \bibinfo {author} {\bibfnamefont {J.}~\bibnamefont
  {Garcia-Ojalvo}},\ and\ \bibinfo {author} {\bibfnamefont {M.~B.}\
  \bibnamefont {Elowitz}},\ }\bibfield  {title} {\bibinfo {title} {Synthetic
  multistability in mammalian cells},\ }\href@noop {} {\bibfield  {journal}
  {\bibinfo  {journal} {Science}\ }\textbf {\bibinfo {volume} {375}},\ \bibinfo
  {pages} {eabg9765} (\bibinfo {year} {2022})}\BibitemShut {NoStop}%
\bibitem [{\citenamefont {Roquet}\ \emph {et~al.}(2016)\citenamefont {Roquet},
  \citenamefont {Soleimany}, \citenamefont {Ferris}, \citenamefont {Aaronson},\
  and\ \citenamefont {Lu}}]{roquet2016synthetic}%
  \BibitemOpen
  \bibfield  {author} {\bibinfo {author} {\bibfnamefont {N.}~\bibnamefont
  {Roquet}}, \bibinfo {author} {\bibfnamefont {A.~P.}\ \bibnamefont
  {Soleimany}}, \bibinfo {author} {\bibfnamefont {A.~C.}\ \bibnamefont
  {Ferris}}, \bibinfo {author} {\bibfnamefont {S.}~\bibnamefont {Aaronson}},\
  and\ \bibinfo {author} {\bibfnamefont {T.~K.}\ \bibnamefont {Lu}},\
  }\bibfield  {title} {\bibinfo {title} {Synthetic recombinase-based state
  machines in living cells},\ }\href@noop {} {\bibfield  {journal} {\bibinfo
  {journal} {Science}\ }\textbf {\bibinfo {volume} {353}},\ \bibinfo {pages}
  {aad8559} (\bibinfo {year} {2016})}\BibitemShut {NoStop}%
\bibitem [{\citenamefont {Qian}\ and\ \citenamefont
  {Winfree}(2011)}]{qian2011scaling}%
  \BibitemOpen
  \bibfield  {author} {\bibinfo {author} {\bibfnamefont {L.}~\bibnamefont
  {Qian}}\ and\ \bibinfo {author} {\bibfnamefont {E.}~\bibnamefont {Winfree}},\
  }\bibfield  {title} {\bibinfo {title} {Scaling up digital circuit computation
  with dna strand displacement cascades},\ }\href@noop {} {\bibfield  {journal}
  {\bibinfo  {journal} {science}\ }\textbf {\bibinfo {volume} {332}},\ \bibinfo
  {pages} {1196} (\bibinfo {year} {2011})}\BibitemShut {NoStop}%
\bibitem [{\citenamefont {Chen}\ and\ \citenamefont
  {Elowitz}(2021)}]{chen2021programmable}%
  \BibitemOpen
  \bibfield  {author} {\bibinfo {author} {\bibfnamefont {Z.}~\bibnamefont
  {Chen}}\ and\ \bibinfo {author} {\bibfnamefont {M.~B.}\ \bibnamefont
  {Elowitz}},\ }\bibfield  {title} {\bibinfo {title} {Programmable protein
  circuit design},\ }\href@noop {} {\bibfield  {journal} {\bibinfo  {journal}
  {Cell}\ }\textbf {\bibinfo {volume} {184}},\ \bibinfo {pages} {2284}
  (\bibinfo {year} {2021})}\BibitemShut {NoStop}%
\bibitem [{\citenamefont {Crick}(1984)}]{crick1984neurobiology}%
  \BibitemOpen
  \bibfield  {author} {\bibinfo {author} {\bibfnamefont {F.}~\bibnamefont
  {Crick}},\ }\bibfield  {title} {\bibinfo {title} {Neurobiology: Memory and
  molecular turnover},\ }\href@noop {} {\bibfield  {journal} {\bibinfo
  {journal} {Nature}\ }\textbf {\bibinfo {volume} {312}},\ \bibinfo {pages}
  {101} (\bibinfo {year} {1984})}\BibitemShut {NoStop}%
\bibitem [{\citenamefont {Hamming}(1950)}]{hamming1950error}%
  \BibitemOpen
  \bibfield  {author} {\bibinfo {author} {\bibfnamefont {R.~W.}\ \bibnamefont
  {Hamming}},\ }\bibfield  {title} {\bibinfo {title} {Error detecting and error
  correcting codes},\ }\href@noop {} {\bibfield  {journal} {\bibinfo  {journal}
  {The Bell system technical journal}\ }\textbf {\bibinfo {volume} {29}},\
  \bibinfo {pages} {147} (\bibinfo {year} {1950})}\BibitemShut {NoStop}%
\bibitem [{\citenamefont {Lisman}(1985)}]{lisman1985mechanism}%
  \BibitemOpen
  \bibfield  {author} {\bibinfo {author} {\bibfnamefont {J.~E.}\ \bibnamefont
  {Lisman}},\ }\bibfield  {title} {\bibinfo {title} {A mechanism for memory
  storage insensitive to molecular turnover: a bistable autophosphorylating
  kinase.},\ }\href@noop {} {\bibfield  {journal} {\bibinfo  {journal}
  {Proceedings of the National Academy of Sciences}\ }\textbf {\bibinfo
  {volume} {82}},\ \bibinfo {pages} {3055} (\bibinfo {year}
  {1985})}\BibitemShut {NoStop}%
\bibitem [{\citenamefont {Stratton}(2022)}]{stratton2022start}%
  \BibitemOpen
  \bibfield  {author} {\bibinfo {author} {\bibfnamefont {M.~M.}\ \bibnamefont
  {Stratton}},\ }\bibfield  {title} {\bibinfo {title} {Start spreading the
  news! camkii shares activity with naive molecules},\ }\href@noop {}
  {\bibfield  {journal} {\bibinfo  {journal} {Proceedings of the National
  Academy of Sciences}\ }\textbf {\bibinfo {volume} {119}},\ \bibinfo {pages}
  {e2216529119} (\bibinfo {year} {2022})}\BibitemShut {NoStop}%
\bibitem [{\citenamefont {Osat}\ and\ \citenamefont
  {Golestanian}(2023)}]{osat2023non}%
  \BibitemOpen
  \bibfield  {author} {\bibinfo {author} {\bibfnamefont {S.}~\bibnamefont
  {Osat}}\ and\ \bibinfo {author} {\bibfnamefont {R.}~\bibnamefont
  {Golestanian}},\ }\bibfield  {title} {\bibinfo {title} {Non-reciprocal
  multifarious self-organization},\ }\href@noop {} {\bibfield  {journal}
  {\bibinfo  {journal} {Nature Nanotechnology}\ }\textbf {\bibinfo {volume}
  {18}},\ \bibinfo {pages} {79} (\bibinfo {year} {2023})}\BibitemShut {NoStop}%
\bibitem [{\citenamefont {Benoist}\ and\ \citenamefont
  {Sartori}(2025)}]{benoist2025high}%
  \BibitemOpen
  \bibfield  {author} {\bibinfo {author} {\bibfnamefont {F.}~\bibnamefont
  {Benoist}}\ and\ \bibinfo {author} {\bibfnamefont {P.}~\bibnamefont
  {Sartori}},\ }\bibfield  {title} {\bibinfo {title} {High-speed combinatorial
  polymerization through kinetic-trap encoding},\ }\href@noop {} {\bibfield
  {journal} {\bibinfo  {journal} {Physical Review Letters}\ }\textbf {\bibinfo
  {volume} {134}},\ \bibinfo {pages} {038402} (\bibinfo {year}
  {2025})}\BibitemShut {NoStop}%
\bibitem [{\citenamefont {Genthon}\ \emph {et~al.}(2025)\citenamefont
  {Genthon}, \citenamefont {Modes}, \citenamefont {J\"ulicher},\ and\
  \citenamefont {Grill}}]{PhysRevLett.134.068402}%
  \BibitemOpen
  \bibfield  {author} {\bibinfo {author} {\bibfnamefont {A.}~\bibnamefont
  {Genthon}}, \bibinfo {author} {\bibfnamefont {C.~D.}\ \bibnamefont {Modes}},
  \bibinfo {author} {\bibfnamefont {F.}~\bibnamefont {J\"ulicher}},\ and\
  \bibinfo {author} {\bibfnamefont {S.~W.}\ \bibnamefont {Grill}},\ }\bibfield
  {title} {\bibinfo {title} {Nonequilibrium transitions in a template copying
  ensemble},\ }\href {https://doi.org/10.1103/PhysRevLett.134.068402}
  {\bibfield  {journal} {\bibinfo  {journal} {Phys. Rev. Lett.}\ }\textbf
  {\bibinfo {volume} {134}},\ \bibinfo {pages} {068402} (\bibinfo {year}
  {2025})}\BibitemShut {NoStop}%
\bibitem [{\citenamefont {Tu}(2008)}]{tu2008nonequilibrium}%
  \BibitemOpen
  \bibfield  {author} {\bibinfo {author} {\bibfnamefont {Y.}~\bibnamefont
  {Tu}},\ }\bibfield  {title} {\bibinfo {title} {The nonequilibrium mechanism
  for ultrasensitivity in a biological switch: sensing by maxwell's demons},\
  }\href@noop {} {\bibfield  {journal} {\bibinfo  {journal} {Proceedings of the
  National Academy of Sciences}\ }\textbf {\bibinfo {volume} {105}},\ \bibinfo
  {pages} {11737} (\bibinfo {year} {2008})}\BibitemShut {NoStop}%
\bibitem [{\citenamefont {Mahdavi}\ \emph {et~al.}(2024)\citenamefont
  {Mahdavi}, \citenamefont {Salmon}, \citenamefont {Daghlian}, \citenamefont
  {Garcia},\ and\ \citenamefont {Phillips}}]{mahdavi2024flexibility}%
  \BibitemOpen
  \bibfield  {author} {\bibinfo {author} {\bibfnamefont {S.~D.}\ \bibnamefont
  {Mahdavi}}, \bibinfo {author} {\bibfnamefont {G.~L.}\ \bibnamefont {Salmon}},
  \bibinfo {author} {\bibfnamefont {P.}~\bibnamefont {Daghlian}}, \bibinfo
  {author} {\bibfnamefont {H.~G.}\ \bibnamefont {Garcia}},\ and\ \bibinfo
  {author} {\bibfnamefont {R.}~\bibnamefont {Phillips}},\ }\bibfield  {title}
  {\bibinfo {title} {Flexibility and sensitivity in gene regulation out of
  equilibrium},\ }\href@noop {} {\bibfield  {journal} {\bibinfo  {journal}
  {Proceedings of the National Academy of Sciences}\ }\textbf {\bibinfo
  {volume} {121}},\ \bibinfo {pages} {e2411395121} (\bibinfo {year}
  {2024})}\BibitemShut {NoStop}%
\bibitem [{\citenamefont {Monod}\ \emph {et~al.}(1965)\citenamefont {Monod},
  \citenamefont {Wyman},\ and\ \citenamefont {Changeux}}]{monod1965nature}%
  \BibitemOpen
  \bibfield  {author} {\bibinfo {author} {\bibfnamefont {J.}~\bibnamefont
  {Monod}}, \bibinfo {author} {\bibfnamefont {J.}~\bibnamefont {Wyman}},\ and\
  \bibinfo {author} {\bibfnamefont {J.-P.}\ \bibnamefont {Changeux}},\
  }\bibfield  {title} {\bibinfo {title} {On the nature of allosteric
  transitions: a plausible model},\ }\href@noop {} {\bibfield  {journal}
  {\bibinfo  {journal} {Journal of molecular biology}\ }\textbf {\bibinfo
  {volume} {12}},\ \bibinfo {pages} {88} (\bibinfo {year} {1965})}\BibitemShut
  {NoStop}%
\bibitem [{\citenamefont {Phillips}(2020)}]{phillips2020molecular}%
  \BibitemOpen
  \bibfield  {author} {\bibinfo {author} {\bibfnamefont {R.}~\bibnamefont
  {Phillips}},\ }\href@noop {} {\emph {\bibinfo {title} {The molecular switch:
  Signaling and Allostery}}}\ (\bibinfo  {publisher} {Princeton University
  Press},\ \bibinfo {year} {2020})\BibitemShut {NoStop}%
\bibitem [{\citenamefont {Rust}\ \emph {et~al.}(2007)\citenamefont {Rust},
  \citenamefont {Markson}, \citenamefont {Lane}, \citenamefont {Fisher},\ and\
  \citenamefont {O'Shea}}]{rust2007ordered}%
  \BibitemOpen
  \bibfield  {author} {\bibinfo {author} {\bibfnamefont {M.~J.}\ \bibnamefont
  {Rust}}, \bibinfo {author} {\bibfnamefont {J.~S.}\ \bibnamefont {Markson}},
  \bibinfo {author} {\bibfnamefont {W.~S.}\ \bibnamefont {Lane}}, \bibinfo
  {author} {\bibfnamefont {D.~S.}\ \bibnamefont {Fisher}},\ and\ \bibinfo
  {author} {\bibfnamefont {E.~K.}\ \bibnamefont {O'Shea}},\ }\bibfield  {title}
  {\bibinfo {title} {Ordered phosphorylation governs oscillation of a
  three-protein circadian clock},\ }\href@noop {} {\bibfield  {journal}
  {\bibinfo  {journal} {Science}\ }\textbf {\bibinfo {volume} {318}},\ \bibinfo
  {pages} {809} (\bibinfo {year} {2007})}\BibitemShut {NoStop}%
\bibitem [{\citenamefont {Murugan}\ and\ \citenamefont
  {Vaikuntanathan}(2017)}]{murugan2017topologically}%
  \BibitemOpen
  \bibfield  {author} {\bibinfo {author} {\bibfnamefont {A.}~\bibnamefont
  {Murugan}}\ and\ \bibinfo {author} {\bibfnamefont {S.}~\bibnamefont
  {Vaikuntanathan}},\ }\bibfield  {title} {\bibinfo {title} {Topologically
  protected modes in non-equilibrium stochastic systems},\ }\href@noop {}
  {\bibfield  {journal} {\bibinfo  {journal} {Nature communications}\ }\textbf
  {\bibinfo {volume} {8}},\ \bibinfo {pages} {13881} (\bibinfo {year}
  {2017})}\BibitemShut {NoStop}%
\bibitem [{\citenamefont {Tang}\ \emph {et~al.}(2021)\citenamefont {Tang},
  \citenamefont {Agudo-Canalejo},\ and\ \citenamefont
  {Golestanian}}]{PhysRevX.11.031015}%
  \BibitemOpen
  \bibfield  {author} {\bibinfo {author} {\bibfnamefont {E.}~\bibnamefont
  {Tang}}, \bibinfo {author} {\bibfnamefont {J.}~\bibnamefont
  {Agudo-Canalejo}},\ and\ \bibinfo {author} {\bibfnamefont {R.}~\bibnamefont
  {Golestanian}},\ }\bibfield  {title} {\bibinfo {title} {Topology protects
  chiral edge currents in stochastic systems},\ }\href
  {https://doi.org/10.1103/PhysRevX.11.031015} {\bibfield  {journal} {\bibinfo
  {journal} {Phys. Rev. X}\ }\textbf {\bibinfo {volume} {11}},\ \bibinfo
  {pages} {031015} (\bibinfo {year} {2021})}\BibitemShut {NoStop}%
\bibitem [{\citenamefont {Agudo-Canalejo}\ and\ \citenamefont
  {Tang}(2025)}]{agudocanalejo2024topologicalphasesdiscretestochastic}%
  \BibitemOpen
  \bibfield  {author} {\bibinfo {author} {\bibfnamefont {J.}~\bibnamefont
  {Agudo-Canalejo}}\ and\ \bibinfo {author} {\bibfnamefont {E.}~\bibnamefont
  {Tang}},\ }\bibfield  {title} {\bibinfo {title} {Topological phases in
  discrete stochastic systems},\ }\href
  {https://doi.org/10.1088/1361-6633/ae07fd} {\bibfield  {journal} {\bibinfo
  {journal} {Rep. Prog. Phys.}\ }\textbf {\bibinfo {volume} {88}},\ \bibinfo
  {pages} {102601} (\bibinfo {year} {2025})}\BibitemShut {NoStop}%
\bibitem [{\citenamefont {Wodak}\ \emph {et~al.}(2019)\citenamefont {Wodak},
  \citenamefont {Paci}, \citenamefont {Dokholyan}, \citenamefont {Berezovsky},
  \citenamefont {Horovitz}, \citenamefont {Li}, \citenamefont {Hilser},
  \citenamefont {Bahar}, \citenamefont {Karanicolas}, \citenamefont {Stock},
  \citenamefont {Hamm}, \citenamefont {Stote}, \citenamefont {Eberhardt},
  \citenamefont {Chebaro}, \citenamefont {Dejaegere}, \citenamefont {Cecchini},
  \citenamefont {Changeux}, \citenamefont {Bolhuis}, \citenamefont {Vreede},
  \citenamefont {Faccioli}, \citenamefont {Orioli}, \citenamefont {Ravasio},
  \citenamefont {Yan}, \citenamefont {Brito}, \citenamefont {Wyart},
  \citenamefont {Gkeka}, \citenamefont {Rivalta}, \citenamefont {Palermo},
  \citenamefont {McCammon}, \citenamefont {Panecka-Hofman}, \citenamefont
  {Wade}, \citenamefont {Di~Pizio}, \citenamefont {Niv}, \citenamefont
  {Nussinov}, \citenamefont {Tsai}, \citenamefont {Jang}, \citenamefont
  {Padhorny}, \citenamefont {Kozakov},\ and\ \citenamefont
  {McLeish}}]{wodak2019allostery}%
  \BibitemOpen
  \bibfield  {author} {\bibinfo {author} {\bibfnamefont {S.~J.}\ \bibnamefont
  {Wodak}}, \bibinfo {author} {\bibfnamefont {E.}~\bibnamefont {Paci}},
  \bibinfo {author} {\bibfnamefont {N.~V.}\ \bibnamefont {Dokholyan}}, \bibinfo
  {author} {\bibfnamefont {I.~N.}\ \bibnamefont {Berezovsky}}, \bibinfo
  {author} {\bibfnamefont {A.}~\bibnamefont {Horovitz}}, \bibinfo {author}
  {\bibfnamefont {J.}~\bibnamefont {Li}}, \bibinfo {author} {\bibfnamefont
  {V.~J.}\ \bibnamefont {Hilser}}, \bibinfo {author} {\bibfnamefont
  {I.}~\bibnamefont {Bahar}}, \bibinfo {author} {\bibfnamefont
  {J.}~\bibnamefont {Karanicolas}}, \bibinfo {author} {\bibfnamefont
  {G.}~\bibnamefont {Stock}}, \bibinfo {author} {\bibfnamefont
  {P.}~\bibnamefont {Hamm}}, \bibinfo {author} {\bibfnamefont {R.~H.}\
  \bibnamefont {Stote}}, \bibinfo {author} {\bibfnamefont {J.}~\bibnamefont
  {Eberhardt}}, \bibinfo {author} {\bibfnamefont {Y.}~\bibnamefont {Chebaro}},
  \bibinfo {author} {\bibfnamefont {A.}~\bibnamefont {Dejaegere}}, \bibinfo
  {author} {\bibfnamefont {M.}~\bibnamefont {Cecchini}}, \bibinfo {author}
  {\bibfnamefont {J.-P.}\ \bibnamefont {Changeux}}, \bibinfo {author}
  {\bibfnamefont {P.~G.}\ \bibnamefont {Bolhuis}}, \bibinfo {author}
  {\bibfnamefont {J.}~\bibnamefont {Vreede}}, \bibinfo {author} {\bibfnamefont
  {P.}~\bibnamefont {Faccioli}}, \bibinfo {author} {\bibfnamefont
  {S.}~\bibnamefont {Orioli}}, \bibinfo {author} {\bibfnamefont
  {R.}~\bibnamefont {Ravasio}}, \bibinfo {author} {\bibfnamefont
  {L.}~\bibnamefont {Yan}}, \bibinfo {author} {\bibfnamefont {C.}~\bibnamefont
  {Brito}}, \bibinfo {author} {\bibfnamefont {M.}~\bibnamefont {Wyart}},
  \bibinfo {author} {\bibfnamefont {P.}~\bibnamefont {Gkeka}}, \bibinfo
  {author} {\bibfnamefont {I.}~\bibnamefont {Rivalta}}, \bibinfo {author}
  {\bibfnamefont {G.}~\bibnamefont {Palermo}}, \bibinfo {author} {\bibfnamefont
  {J.~A.}\ \bibnamefont {McCammon}}, \bibinfo {author} {\bibfnamefont
  {J.}~\bibnamefont {Panecka-Hofman}}, \bibinfo {author} {\bibfnamefont
  {R.~C.}\ \bibnamefont {Wade}}, \bibinfo {author} {\bibfnamefont
  {A.}~\bibnamefont {Di~Pizio}}, \bibinfo {author} {\bibfnamefont {M.~Y.}\
  \bibnamefont {Niv}}, \bibinfo {author} {\bibfnamefont {R.}~\bibnamefont
  {Nussinov}}, \bibinfo {author} {\bibfnamefont {C.-J.}\ \bibnamefont {Tsai}},
  \bibinfo {author} {\bibfnamefont {H.}~\bibnamefont {Jang}}, \bibinfo {author}
  {\bibfnamefont {D.}~\bibnamefont {Padhorny}}, \bibinfo {author}
  {\bibfnamefont {D.}~\bibnamefont {Kozakov}},\ and\ \bibinfo {author}
  {\bibfnamefont {T.}~\bibnamefont {McLeish}},\ }\bibfield  {title} {\bibinfo
  {title} {Allostery in its many disguises: From theory to applications},\
  }\href {https://doi.org/10.1016/j.str.2019.01.003} {\bibfield  {journal}
  {\bibinfo  {journal} {Structure}\ }\textbf {\bibinfo {volume} {27}},\
  \bibinfo {pages} {566–578} (\bibinfo {year} {2019})}\BibitemShut {NoStop}%
\bibitem [{\citenamefont {Tomita}\ \emph {et~al.}(2005)\citenamefont {Tomita},
  \citenamefont {Nakajima}, \citenamefont {Kondo},\ and\ \citenamefont
  {Iwasaki}}]{tomita2005no}%
  \BibitemOpen
  \bibfield  {author} {\bibinfo {author} {\bibfnamefont {J.}~\bibnamefont
  {Tomita}}, \bibinfo {author} {\bibfnamefont {M.}~\bibnamefont {Nakajima}},
  \bibinfo {author} {\bibfnamefont {T.}~\bibnamefont {Kondo}},\ and\ \bibinfo
  {author} {\bibfnamefont {H.}~\bibnamefont {Iwasaki}},\ }\bibfield  {title}
  {\bibinfo {title} {No transcription-translation feedback in circadian rhythm
  of kaic phosphorylation},\ }\href@noop {} {\bibfield  {journal} {\bibinfo
  {journal} {Science}\ }\textbf {\bibinfo {volume} {307}},\ \bibinfo {pages}
  {251} (\bibinfo {year} {2005})}\BibitemShut {NoStop}%
\bibitem [{\citenamefont {Stadtman}\ and\ \citenamefont
  {Chock}(1977)}]{stadtman1977superiority}%
  \BibitemOpen
  \bibfield  {author} {\bibinfo {author} {\bibfnamefont {E.}~\bibnamefont
  {Stadtman}}\ and\ \bibinfo {author} {\bibfnamefont {P.}~\bibnamefont
  {Chock}},\ }\bibfield  {title} {\bibinfo {title} {Superiority of
  interconvertible enzyme cascades in metabolic regulation: analysis of
  monocyclic systems.},\ }\href@noop {} {\bibfield  {journal} {\bibinfo
  {journal} {Proceedings of the National Academy of Sciences}\ }\textbf
  {\bibinfo {volume} {74}},\ \bibinfo {pages} {2761} (\bibinfo {year}
  {1977})}\BibitemShut {NoStop}%
\bibitem [{\citenamefont {Detwiler}\ \emph {et~al.}(2000)\citenamefont
  {Detwiler}, \citenamefont {Ramanathan}, \citenamefont {Sengupta},\ and\
  \citenamefont {Shraiman}}]{detwiler2000engineering}%
  \BibitemOpen
  \bibfield  {author} {\bibinfo {author} {\bibfnamefont {P.~B.}\ \bibnamefont
  {Detwiler}}, \bibinfo {author} {\bibfnamefont {S.}~\bibnamefont
  {Ramanathan}}, \bibinfo {author} {\bibfnamefont {A.}~\bibnamefont
  {Sengupta}},\ and\ \bibinfo {author} {\bibfnamefont {B.~I.}\ \bibnamefont
  {Shraiman}},\ }\bibfield  {title} {\bibinfo {title} {Engineering aspects of
  enzymatic signal transduction: photoreceptors in the retina},\ }\href@noop {}
  {\bibfield  {journal} {\bibinfo  {journal} {Biophysical journal}\ }\textbf
  {\bibinfo {volume} {79}},\ \bibinfo {pages} {2801} (\bibinfo {year}
  {2000})}\BibitemShut {NoStop}%
\bibitem [{\citenamefont {Wolfram}(1983)}]{RevModPhys.55.601}%
  \BibitemOpen
  \bibfield  {author} {\bibinfo {author} {\bibfnamefont {S.}~\bibnamefont
  {Wolfram}},\ }\bibfield  {title} {\bibinfo {title} {Statistical mechanics of
  cellular automata},\ }\href {https://doi.org/10.1103/RevModPhys.55.601}
  {\bibfield  {journal} {\bibinfo  {journal} {Rev. Mod. Phys.}\ }\textbf
  {\bibinfo {volume} {55}},\ \bibinfo {pages} {601} (\bibinfo {year}
  {1983})}\BibitemShut {NoStop}%
\bibitem [{\citenamefont {Cornforth}\ \emph {et~al.}(2005)\citenamefont
  {Cornforth}, \citenamefont {Green},\ and\ \citenamefont
  {Newth}}]{cornforth2005ordered}%
  \BibitemOpen
  \bibfield  {author} {\bibinfo {author} {\bibfnamefont {D.}~\bibnamefont
  {Cornforth}}, \bibinfo {author} {\bibfnamefont {D.~G.}\ \bibnamefont
  {Green}},\ and\ \bibinfo {author} {\bibfnamefont {D.}~\bibnamefont {Newth}},\
  }\bibfield  {title} {\bibinfo {title} {Ordered asynchronous processes in
  multi-agent systems},\ }\href@noop {} {\bibfield  {journal} {\bibinfo
  {journal} {Physica D: Nonlinear Phenomena}\ }\textbf {\bibinfo {volume}
  {204}},\ \bibinfo {pages} {70} (\bibinfo {year} {2005})}\BibitemShut
  {NoStop}%
\bibitem [{\citenamefont {Li}\ and\ \citenamefont
  {Packard}(1990)}]{li1990structure}%
  \BibitemOpen
  \bibfield  {author} {\bibinfo {author} {\bibfnamefont {W.}~\bibnamefont
  {Li}}\ and\ \bibinfo {author} {\bibfnamefont {N.}~\bibnamefont {Packard}},\
  }\bibfield  {title} {\bibinfo {title} {The structure of the elementary
  cellular automata rule space},\ }\href@noop {} {\bibfield  {journal}
  {\bibinfo  {journal} {Complex systems}\ }\textbf {\bibinfo {volume} {4}},\
  \bibinfo {pages} {281} (\bibinfo {year} {1990})}\BibitemShut {NoStop}%
\bibitem [{sup()}]{suppmat}%
  \BibitemOpen
  \href@noop {} {}\bibinfo {note} {See Supplemental Material [url] for
  additional figures demonstrating how different features vary with the complex
  size $N$ (Figs. S1 and S4--S8), and the error-correcting capabilities of all
  four error-correcting rules (Figs. S2 and S3).}\BibitemShut {Stop}%
\bibitem [{\citenamefont {Hopcroft}\ \emph {et~al.}(2001)\citenamefont
  {Hopcroft}, \citenamefont {Motwani},\ and\ \citenamefont
  {Ullman}}]{hopcroft2001introduction}%
  \BibitemOpen
  \bibfield  {author} {\bibinfo {author} {\bibfnamefont {J.~E.}\ \bibnamefont
  {Hopcroft}}, \bibinfo {author} {\bibfnamefont {R.}~\bibnamefont {Motwani}},\
  and\ \bibinfo {author} {\bibfnamefont {J.~D.}\ \bibnamefont {Ullman}},\
  }\bibfield  {title} {\bibinfo {title} {Introduction to automata theory,
  languages, and computation},\ }\href@noop {} {\bibfield  {journal} {\bibinfo
  {journal} {Acm Sigact News}\ }\textbf {\bibinfo {volume} {32}},\ \bibinfo
  {pages} {60} (\bibinfo {year} {2001})}\BibitemShut {NoStop}%
\bibitem [{\citenamefont {Ito}\ \emph {et~al.}(2007)\citenamefont {Ito},
  \citenamefont {Kageyama}, \citenamefont {Mutsuda}, \citenamefont {Nakajima},
  \citenamefont {Oyama},\ and\ \citenamefont {Kondo}}]{ito2007autonomous}%
  \BibitemOpen
  \bibfield  {author} {\bibinfo {author} {\bibfnamefont {H.}~\bibnamefont
  {Ito}}, \bibinfo {author} {\bibfnamefont {H.}~\bibnamefont {Kageyama}},
  \bibinfo {author} {\bibfnamefont {M.}~\bibnamefont {Mutsuda}}, \bibinfo
  {author} {\bibfnamefont {M.}~\bibnamefont {Nakajima}}, \bibinfo {author}
  {\bibfnamefont {T.}~\bibnamefont {Oyama}},\ and\ \bibinfo {author}
  {\bibfnamefont {T.}~\bibnamefont {Kondo}},\ }\bibfield  {title} {\bibinfo
  {title} {Autonomous synchronization of the circadian kaic phosphorylation
  rhythm},\ }\href@noop {} {\bibfield  {journal} {\bibinfo  {journal} {Nature
  structural \& molecular biology}\ }\textbf {\bibinfo {volume} {14}},\
  \bibinfo {pages} {1084} (\bibinfo {year} {2007})}\BibitemShut {NoStop}%
\bibitem [{\citenamefont {Kuhlman}\ \emph {et~al.}(2003)\citenamefont
  {Kuhlman}, \citenamefont {Dantas}, \citenamefont {Ireton}, \citenamefont
  {Varani}, \citenamefont {Stoddard},\ and\ \citenamefont
  {Baker}}]{kuhlman2003design}%
  \BibitemOpen
  \bibfield  {author} {\bibinfo {author} {\bibfnamefont {B.}~\bibnamefont
  {Kuhlman}}, \bibinfo {author} {\bibfnamefont {G.}~\bibnamefont {Dantas}},
  \bibinfo {author} {\bibfnamefont {G.~C.}\ \bibnamefont {Ireton}}, \bibinfo
  {author} {\bibfnamefont {G.}~\bibnamefont {Varani}}, \bibinfo {author}
  {\bibfnamefont {B.~L.}\ \bibnamefont {Stoddard}},\ and\ \bibinfo {author}
  {\bibfnamefont {D.}~\bibnamefont {Baker}},\ }\bibfield  {title} {\bibinfo
  {title} {Design of a novel globular protein fold with atomic-level
  accuracy},\ }\href@noop {} {\bibfield  {journal} {\bibinfo  {journal}
  {science}\ }\textbf {\bibinfo {volume} {302}},\ \bibinfo {pages} {1364}
  (\bibinfo {year} {2003})}\BibitemShut {NoStop}%
\bibitem [{\citenamefont {King}\ \emph {et~al.}(2014)\citenamefont {King},
  \citenamefont {Bale}, \citenamefont {Sheffler}, \citenamefont {McNamara},
  \citenamefont {Gonen}, \citenamefont {Gonen}, \citenamefont {Yeates},\ and\
  \citenamefont {Baker}}]{king2014accurate}%
  \BibitemOpen
  \bibfield  {author} {\bibinfo {author} {\bibfnamefont {N.~P.}\ \bibnamefont
  {King}}, \bibinfo {author} {\bibfnamefont {J.~B.}\ \bibnamefont {Bale}},
  \bibinfo {author} {\bibfnamefont {W.}~\bibnamefont {Sheffler}}, \bibinfo
  {author} {\bibfnamefont {D.~E.}\ \bibnamefont {McNamara}}, \bibinfo {author}
  {\bibfnamefont {S.}~\bibnamefont {Gonen}}, \bibinfo {author} {\bibfnamefont
  {T.}~\bibnamefont {Gonen}}, \bibinfo {author} {\bibfnamefont {T.~O.}\
  \bibnamefont {Yeates}},\ and\ \bibinfo {author} {\bibfnamefont
  {D.}~\bibnamefont {Baker}},\ }\bibfield  {title} {\bibinfo {title} {Accurate
  design of co-assembling multi-component protein nanomaterials},\ }\href@noop
  {} {\bibfield  {journal} {\bibinfo  {journal} {Nature}\ }\textbf {\bibinfo
  {volume} {510}},\ \bibinfo {pages} {103} (\bibinfo {year}
  {2014})}\BibitemShut {NoStop}%
\bibitem [{\citenamefont {Pillai}\ \emph {et~al.}(2024)\citenamefont {Pillai},
  \citenamefont {Idris}, \citenamefont {Philomin}, \citenamefont {Weidle},
  \citenamefont {Skotheim}, \citenamefont {Leung}, \citenamefont {Broerman},
  \citenamefont {Demakis}, \citenamefont {Borst}, \citenamefont {Praetorius},\
  and\ \citenamefont {Baker}}]{pillai2024novo}%
  \BibitemOpen
  \bibfield  {author} {\bibinfo {author} {\bibfnamefont {A.}~\bibnamefont
  {Pillai}}, \bibinfo {author} {\bibfnamefont {A.}~\bibnamefont {Idris}},
  \bibinfo {author} {\bibfnamefont {A.}~\bibnamefont {Philomin}}, \bibinfo
  {author} {\bibfnamefont {C.}~\bibnamefont {Weidle}}, \bibinfo {author}
  {\bibfnamefont {R.}~\bibnamefont {Skotheim}}, \bibinfo {author}
  {\bibfnamefont {P.~J.~Y.}\ \bibnamefont {Leung}}, \bibinfo {author}
  {\bibfnamefont {A.}~\bibnamefont {Broerman}}, \bibinfo {author}
  {\bibfnamefont {C.}~\bibnamefont {Demakis}}, \bibinfo {author} {\bibfnamefont
  {A.~J.}\ \bibnamefont {Borst}}, \bibinfo {author} {\bibfnamefont
  {F.}~\bibnamefont {Praetorius}},\ and\ \bibinfo {author} {\bibfnamefont
  {D.}~\bibnamefont {Baker}},\ }\bibfield  {title} {\bibinfo {title} {De novo
  design of allosterically switchable protein assemblies},\ }\href
  {https://doi.org/10.1038/s41586-024-07813-2} {\bibfield  {journal} {\bibinfo
  {journal} {Nature}\ }\textbf {\bibinfo {volume} {632}},\ \bibinfo {pages}
  {911–920} (\bibinfo {year} {2024})}\BibitemShut {NoStop}%
\bibitem [{\citenamefont {Kocka}\ \emph {et~al.}(2026)\citenamefont {Kocka},
  \citenamefont {Husain},\ and\ \citenamefont {Agudo-Canalejo}}]{Kocka2026}%
  \BibitemOpen
  \bibfield  {author} {\bibinfo {author} {\bibfnamefont {J.}~\bibnamefont
  {Kocka}}, \bibinfo {author} {\bibfnamefont {K.}~\bibnamefont {Husain}},\ and\
  \bibinfo {author} {\bibfnamefont {J.}~\bibnamefont {Agudo-Canalejo}},\
  }\bibfield  {title} {\bibinfo {title} {{Code and data for: Nonequilibrium
  protein complexes as molecular automata}}\ }\href
  {https://doi.org/10.5522/04/31497577.v1} {10.5522/04/31497577.v1} (\bibinfo
  {year} {2026})\BibitemShut {NoStop}%
\bibitem [{\citenamefont {Gillespie}(2007)}]{gillespie2007stochastic}%
  \BibitemOpen
  \bibfield  {author} {\bibinfo {author} {\bibfnamefont {D.~T.}\ \bibnamefont
  {Gillespie}},\ }\bibfield  {title} {\bibinfo {title} {Stochastic simulation
  of chemical kinetics},\ }\href@noop {} {\bibfield  {journal} {\bibinfo
  {journal} {Annu. Rev. Phys. Chem.}\ }\textbf {\bibinfo {volume} {58}},\
  \bibinfo {pages} {35} (\bibinfo {year} {2007})}\BibitemShut {NoStop}%
\end{thebibliography}%


\begin{thebibliography}{0}%
\makeatletter
\providecommand \@ifxundefined [1]{%
 \@ifx{#1\undefined}
}%
\providecommand \@ifnum [1]{%
 \ifnum #1\expandafter \@firstoftwo
 \else \expandafter \@secondoftwo
 \fi
}%
\providecommand \@ifx [1]{%
 \ifx #1\expandafter \@firstoftwo
 \else \expandafter \@secondoftwo
 \fi
}%
\providecommand \natexlab [1]{#1}%
\providecommand \enquote  [1]{``#1''}%
\providecommand \bibnamefont  [1]{#1}%
\providecommand \bibfnamefont [1]{#1}%
\providecommand \citenamefont [1]{#1}%
\providecommand \href@noop [0]{\@secondoftwo}%
\providecommand \href [0]{\begingroup \@sanitize@url \@href}%
\providecommand \@href[1]{\@@startlink{#1}\@@href}%
\providecommand \@@href[1]{\endgroup#1\@@endlink}%
\providecommand \@sanitize@url [0]{\catcode `\\12\catcode `\$12\catcode
  `\&12\catcode `\#12\catcode `\^12\catcode `\_12\catcode `\%12\relax}%
\providecommand \@@startlink[1]{}%
\providecommand \@@endlink[0]{}%
\providecommand \url  [0]{\begingroup\@sanitize@url \@url }%
\providecommand \@url [1]{\endgroup\@href {#1}{\urlprefix }}%
\providecommand \urlprefix  [0]{URL }%
\providecommand \Eprint [0]{\href }%
\providecommand \doibase [0]{https://doi.org/}%
\providecommand \selectlanguage [0]{\@gobble}%
\providecommand \bibinfo  [0]{\@secondoftwo}%
\providecommand \bibfield  [0]{\@secondoftwo}%
\providecommand \translation [1]{[#1]}%
\providecommand \BibitemOpen [0]{}%
\providecommand \bibitemStop [0]{}%
\providecommand \bibitemNoStop [0]{.\EOS\space}%
\providecommand \EOS [0]{\spacefactor3000\relax}%
\providecommand \BibitemShut  [1]{\csname bibitem#1\endcsname}%
\let\auto@bib@innerbib\@empty
%</preamble>
\end{thebibliography}%

\end{document}

\title{Supplemental Material\\Nonequilibrium protein complexes as molecular automata}
\author{Jan Kocka}
\address{Department of Physics and Astronomy, University College London, London WC1E 6BT, United Kingdom}
\author{Kabir Husain}
\thanks{These authors contributed equally. Email: kabir.husain@ucl.ac.uk, j.agudo-canalejo@ucl.ac.uk}
\address{Department of Physics and Astronomy, University College London, London WC1E 6BT, United Kingdom}
\address{Laboratory for Molecular Cell Biology, University College London, London WC1E 6BT, United Kingdom}
\author{Jaime Agudo-Canalejo}
\thanks{These authors contributed equally. Email: kabir.husain@ucl.ac.uk, j.agudo-canalejo@ucl.ac.uk}
\address{Department of Physics and Astronomy, University College London, London WC1E 6BT, United Kingdom}

\maketitle

\onecolumngrid

\renewcommand{\thefigure}{S\arabic{figure}}
\renewcommand{\thetable}{S\arabic{table}}

\begin{figure}[H]
    \centering
    \includegraphics[scale=1]{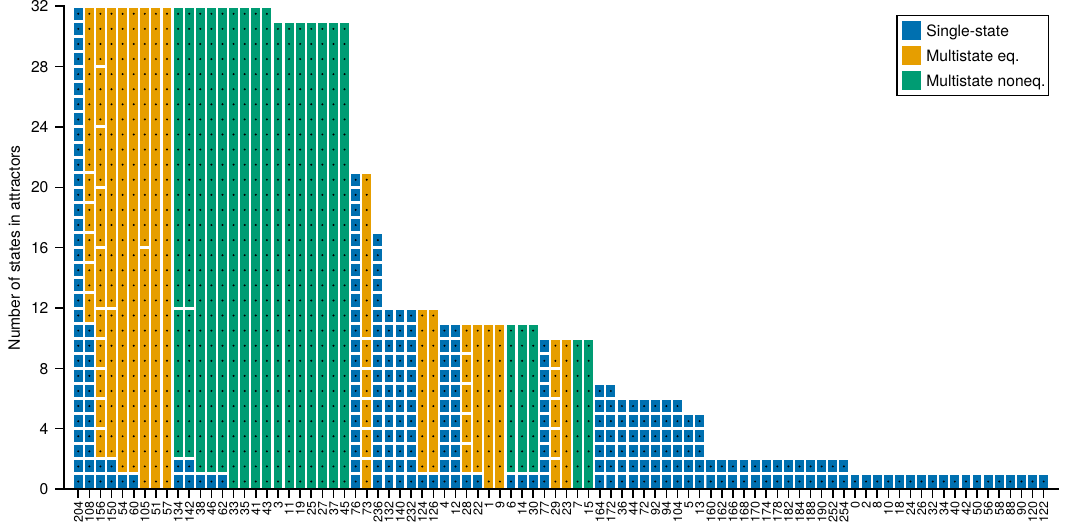}
    \caption{Shown are the number of states in attractors for each of the 88 non-redundant rule-sets (labelled by their Wolfram codes), for a complex of size $N=5$. States in the same attractor are grouped together into boxes, which are coloured to denote the type of attractor. 
    }
\end{figure}

\begin{figure}[H]
    \centering
    \includegraphics[scale=1]{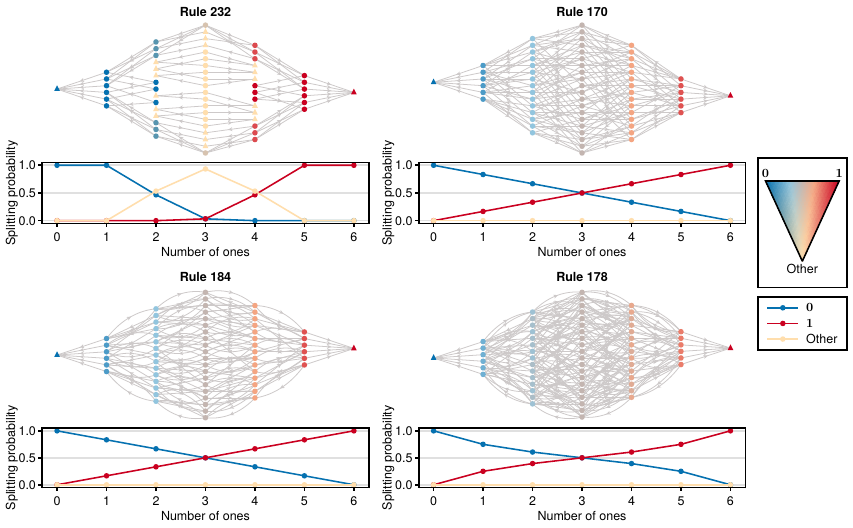}
    \caption{Splitting probability towards the $\textbf{0}$ and $\textbf{1}$ attractors, as well as towards any other attractor, for the four rules that have these two states as attractors, at least one incoming transition into each of them, and are moreover invariant to the bit-flipping operation (rules 170, 178, and 232) or invariant to the combined bit-flipping and mirror symmetry operations (rule 184).
    }
\end{figure}

\begin{figure}[H]
    \centering
    \includegraphics[scale=1]{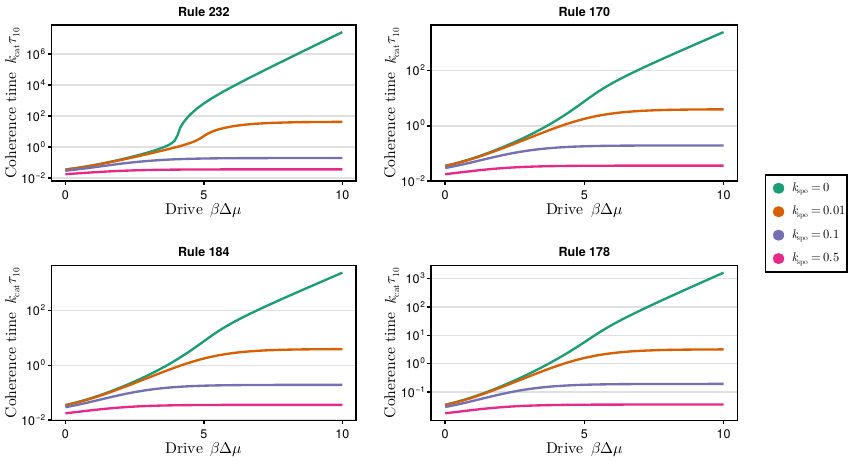}
    \caption{
     Coherence time $\tau_{10}$, representing how long it takes $10\%$ of complexes to exit the $\textbf{0}$ state, as a function of the nonequilibrium driving $\Delta \mu$, for several values of the spontaneous transition rate $k_\mathrm{spo}$ (in units of $k_\mathrm{cat}$). Plots are analogous to those of Fig.~5(b), but here for all four error-correcting rules 232, 170, 184, 178. Rule 232 outperforms the other three, which perform similar to each other. In particular, rules 170 and 184, which have identical splitting probabilities (see Fig.~S2), also have identical coherence times.
    }
\end{figure}

\begin{figure}[H]
    \centering
    \includegraphics[scale=1]{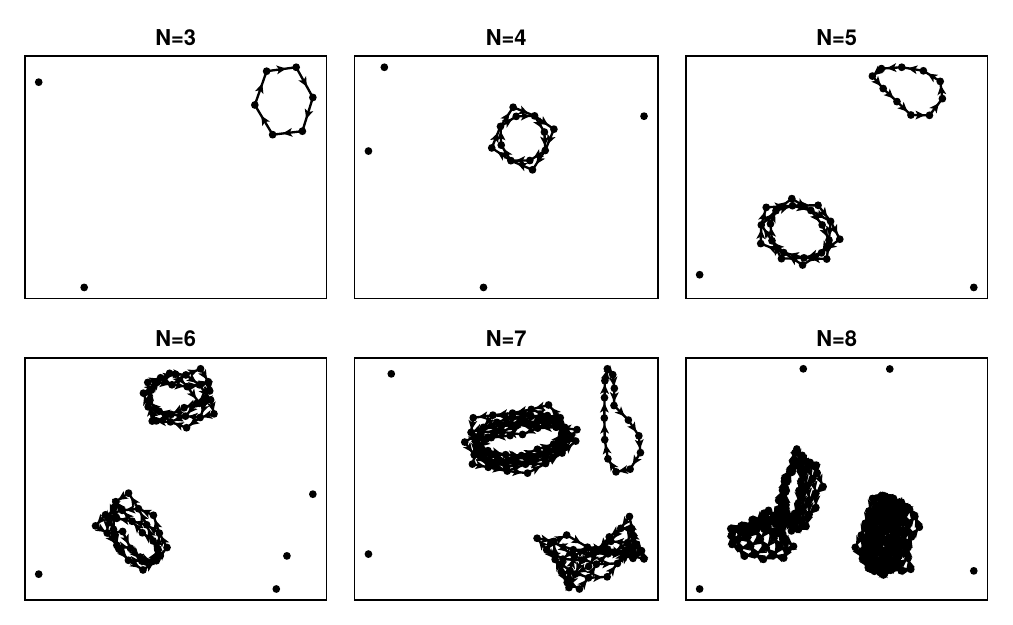}
    \caption{
        Transition graphs for rule 142 for $N$ from 3 to 8.
        We find that there exist nonequilibrium directed loops for all $N$, even or odd. However, there are some parity effects as well: for odd $N$, there is always one unbranched loop (besides other branched ones), and two single-state attractors. For even $N$, all loops are branched, and there are four single-state attractors.
    }
\end{figure}

\begin{figure}[H]
    \centering
    \includegraphics[scale=1]{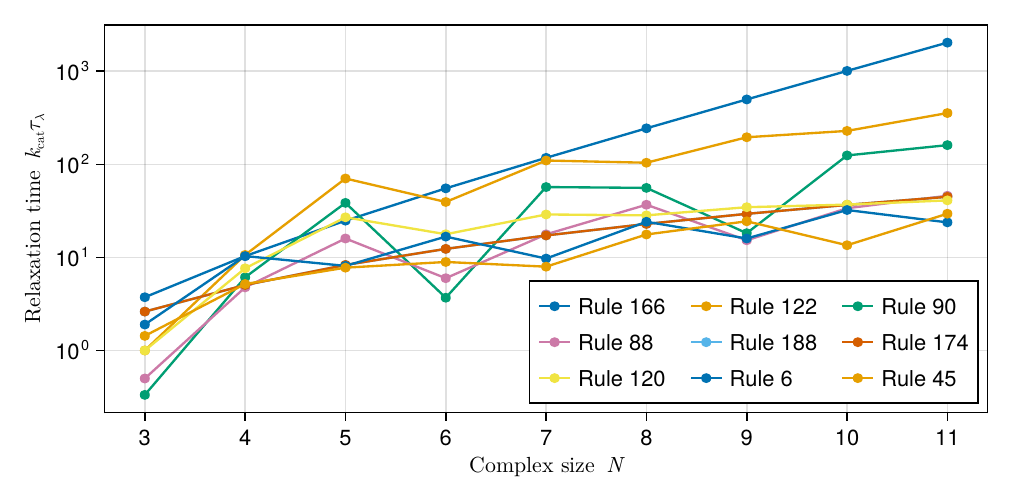}
    \caption{
        Longest relaxation times (inverse spectral gaps), as in Fig.~4(a), as a function of $N$ for the 9 rules with the largest relaxation times (see Fig.~4(a)).
        Rule 166 appears to have the largest relaxation times for $N \ge 6$, and displays the most regular trend with nearly exponential scaling.
        Rules other than 166 appear to have significant parity effects, as evidenced by the alternately increasing and decreasing relaxation times.
    }
\end{figure}

\begin{figure}[H]
    \centering
    \includegraphics[scale=1]{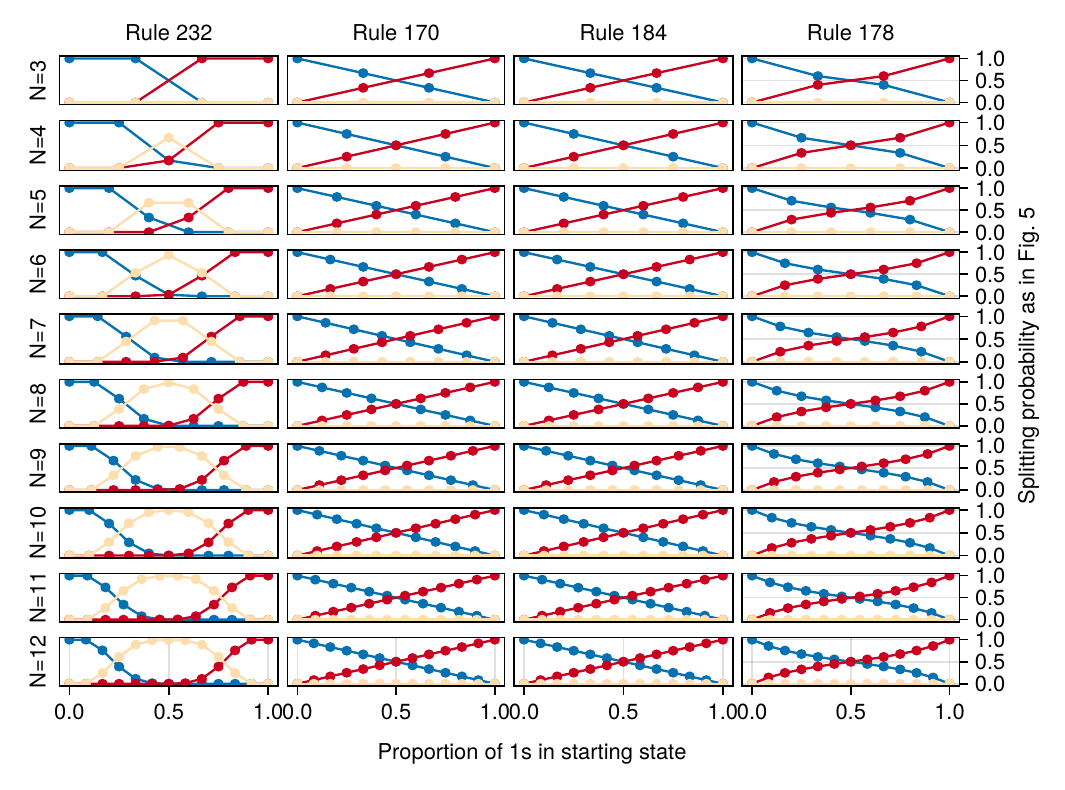}
    \caption{
    Splitting probabilities towards the \textbf{0} and \textbf{1} attractors, as in Fig.~5(a), for all four error-correcting rules and increasing complex sizes $N$.
    The error-correcting capabilities of each rule appear qualitatively unchanged with increasing $N$.
    }
\end{figure}

\begin{figure}[H]
    \centering
    \includegraphics[scale=1]{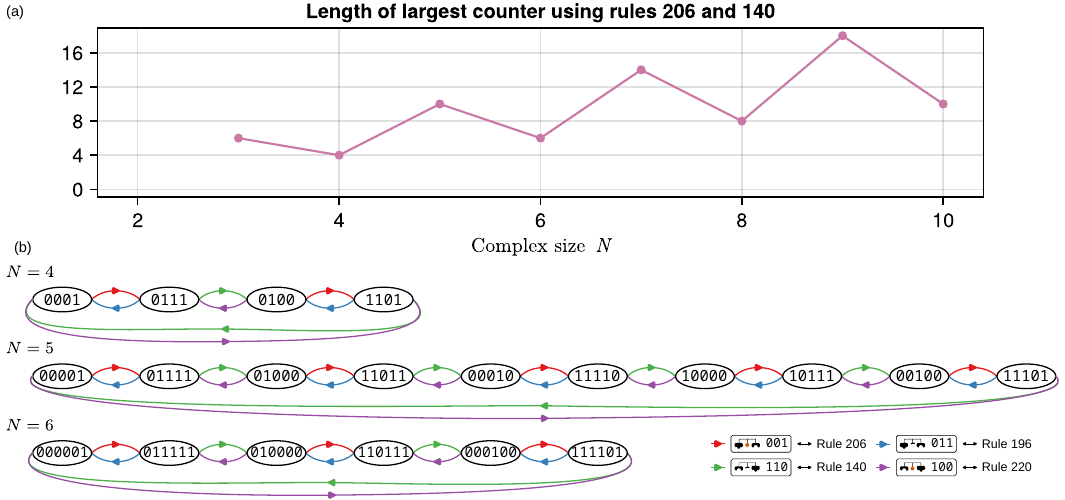}
    \caption{
    (a) Length of the longest counter module realized with rules 206 and 140 (as well as 196 and 220 for backwards counting), as in Fig.~6(b), as a function of $N$.
    If $N$ is even, the maximal length of the counter is  $N$, while if $N$ is odd, the maximal length is $2N$.
    (b) Detailed views of the counter module for $N=4,5,6$, illustrating how the mechanism works and scales with $N$, and highlighting why odd and even $N$ behave differently: for odd $N$, after $N$ steps, the complex returns not to the original state but to its complement. It thus takes another $N$ steps, for a total of $2N$ steps, for the complex to return to the original state.
    }
\end{figure}

\begin{figure}[H]
    \centering
    \includegraphics[scale=1]{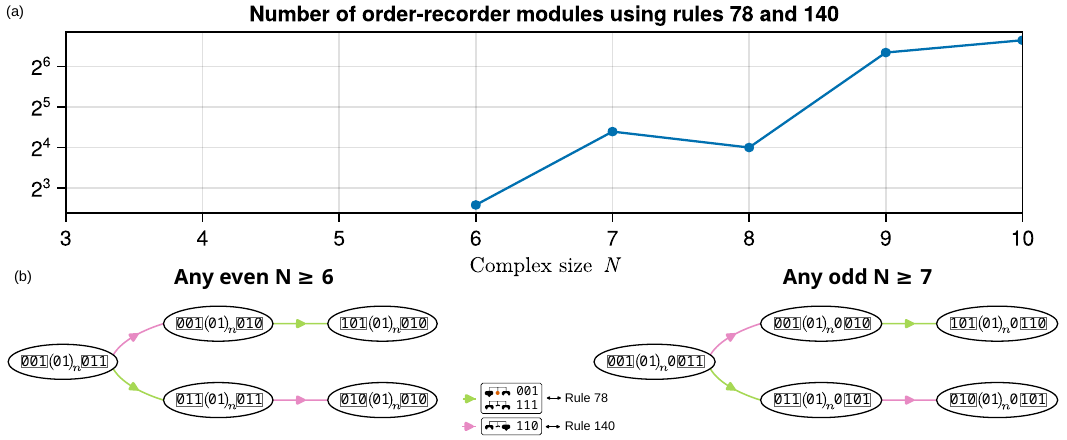}
    \caption{
    (a) Number of order-recorder modules using rules 78 and 140, as in Fig.~6(c), present in the full finite-state machine network as a function of complex size $N$.
    This shows that there are no such modules for $N<6$, and that their number shows odd-even parity effects.
    (b) An example of a pattern that can be used to build order-recorder modules using rules 78 and 140 for any $N$. For $N$ even, we have $n=(N-6)/2$ repetitions of 01. For $N$ odd, we have $n=(N-7)/2$ repetitions.
    Note that there are other possible patterns that work, hence the large number of modules shown in (a).
    }
\end{figure}